\documentclass[aps,pra,preprint,12pt]{revtex4}

\usepackage{amssymb}
\usepackage{amsmath}
\usepackage{bm}
\usepackage{natbib}
\usepackage{graphics,graphicx}
\usepackage{epsfig}
\usepackage{latexsym}
\usepackage{epic,eepic,graphicx,amssymb,amsmath,indentfirst}
\usepackage{bm}
\usepackage{graphicx}
\usepackage[utf8]{inputenc}
\usepackage[T1]{fontenc}
\usepackage{natbib}

\usepackage{multirow}

\usepackage{color}
\usepackage[colorlinks={true}]{hyperref}
\hypersetup{citecolor={blue}, filecolor={blue}, linkcolor={blue}, urlcolor={blue}}
\newcommand{\bra}[1]{\langle #1 \vert}
\newcommand{\ket}[1]{\vert #1 \rangle}

\begin{document}

\title{\bf Simultaneous control of populations and coherence phase of open two-level quantum systems with a single pulse}

\author{Gustavo Fernandes da Costa}
\author{Emanuel Fernandes de Lima}
\email{eflima@ufscar.br}
\affiliation{Departamento de F\'isica, Universidade Federal de S\~ao Carlos (UFSCar)\\ S\~ao Carlos, SP 13565-905, Brazil}
\date{\today}

\begin{abstract}
We address the control of the dynamics of both population and coherence phase in an open two-level quantum system  employing a single external control field. The system dynamics is described by a Markovian master equation that takes into account dephasing and thermal noise. The control is engineered by inverting the underlying equations of motion, which yields an analytical expression for the control field in terms of user-specified time-dependent functions for the population and coherence phase. Our approach allows to dictate not only the initial and final populations and phases, but the full dynamics of these quantities. The chosen functions for population and phase have to conform to certain constraints indicated in our analysis. Our methodology also reveals the possible transitions for given initial conditions and environmental noise parameters.  
\end{abstract}

\maketitle
\newpage

\section{Introduction}

Open two-level quantum systems play a fundamental role in many areas such as Atomic and Molecular Physics and Condensed Matter, standing for an excellent approximation in many practical situations \cite{RevModPhys.59.1,eberly,doi:10.1126/science.1116955,PhysRevA.110.042618,weiss2008quantum}. These systems are of particular importance in quantum technologies, where they serve as qubits, the elementary unit of quantum information, interacting with external environments \cite{Bartee2025,Murch2013}. Analytical solutions, whether exact or derived within the rotating wave approximation (RWA), are especially valuable in control problems, as they yield critical insights into system dynamics and inform the development of effective control strategies.


The prime control problem is to drive a given initial state of the two-level system to an arbitrary target state at some finite time \cite{Brif_2010}. For pure state systems under unitary evolution any target state can be reached if given enough time and with reasonable assumptions on the control fields \cite{10.1063/1.3589404,SCHIRMER2003281}. The situation is more complex in the presence of the environment, where the set of the reachable final states is comparatively smaller. Markovian dynamics is an extreme case where dissipation and relaxation processes irreversibly drive the system to undesired equilibrium states, which cannot be completely overcome by a coherent control field \cite{PhysRevA.70.062321}. Nevertheless, satisfactory control can often be accomplished for some objectives \cite{PhysRevA.88.033420}. Understanding the fine details of the final reachable states is crucial for defining appropriate control goals.  



There are several means of obtaining the solution to a control problem involving closed and open quantum systems \cite{Koch2022,Koch_2016}. Quantum optimal control (QOC) is the prevalent approach, which allows one to obtain the control solutions when the controllability conditions are satisfied \cite{Ansel_2024,Fernandes_2023,10.1063/1.458438,PhysRevLett.125.250403,doi:https://doi.org/10.1002/9783527639700.ch5,10.1063/1.467202}. The sufficient conditions for optimality are expressed as a set of differential equations that must be solved self-consistently by some numerical iterative algorithm. QOC is usually carried out without specifying the underlying dynamics which links the initial to the desired target state. The associated numerical solution are often computationally expensive to obtain and hard to be direct implemented in the laboratory due to its complexity.

Differently from OCT, quantum tracking control (sometimes referred to as inverse quantum control) imposes the system to follow a prescribed dynamics and seeks the control field that generate it \cite{Mirrahimi2005,PhysRevA.98.043429,CHEN19971617,PhysRevA.47.4593}. This is done, for instance, by setting the desired dynamical function for the expectation value of the target observable and inverting the underlying equation of motion in terms of the control field. This procedure is computationally appealing since it avoids heavy iterative schemes of OCT. However, although conceptually simple, this approach often leads to controls that are cursed by unwanted singularities. Much recent effort have been made to circumvent the singularity problem \cite{PhysRevA.108.033106}.

Similarly to quantum tracking, the reverse engineering techniques also seek to build either control fields or Hamiltonians to follow given dynamical constraints \cite{Zhang2017,PhysRevApplied.8.054008,Wang2024,Vitanov_2015}. Concerning the problem of transferring a given initial state to an arbitrary final state with a single control field, reverse engineering methods have been successfully applied to control closed two-level systems  \cite{articleGolubev}. Analytical forms for the field have been designed such that the populations of the system follow a prescribed dynamics. The methodology has been extended to control not only the population but also the phase of the quantum coefficients by means of a single chirped pulse \cite{articleAndras}. The analytical formula for the field is obtained once an integral involving the chosen dynamical functions is performed. In a recent work, an analytical form for the field to control both population and relative phase has been derived in which the prescribed dynamical functions avoid any singularity with no need of calculating integrals \cite{PhysRevA.110.022201}.

The open version of this approach with effects of dephasing and thermal noise taking into account has also been considered \cite{Ran:20,XIAO2021167957,PhysRevA.100.012103,PhysRevA.101.023822}. In \cite{PhysRevA.100.012103}, the authors have considered the control of the populations assuming the phase of the coherence constant. The obtained expression of the control field depends on the solution a differential equation for the absolute value of the coherence. For some fixed initial population, the accessible transitions were determined and the existence of steady states were shown. In \cite{PhysRevA.101.023822}, the time dependent of both populations and the coherence phase has been considered through the representation of the dynamical equations in term of the components of the Bloch vector. In this approach, the generated control fields depended on the calculation of an integral while the control of the coherence phase were not directly accessed through the chosen dynamical functions. 

In the present work, we extend previous investigations \cite{PhysRevA.100.012103,PhysRevA.101.023822} which consider the control of a two-level system under the effects of dephasing and thermal noise including explicitly the control of the coherence phase. The simultaneous control of the populations and of the phase of the coherence is done by a single control pulse. The analytical form of the pulse depends on the chosen dynamical functions for the population and for the phase coherence, which must satisfy some constraints to be valid. The absolute value of the coherence is determined from a differential equation, which can be solved analytic once the population dynamics is specified. Finally, we examine the reachable transitions as a function of the rate of decoherence and thermal noise.

\section{Design of the control field}

We envisage a two-level quantum system under the action of a time-dependent external field during the time interval from $t=0$ to $t=t_f$, following closely previous formulations \cite{PhysRevA.100.012103,PhysRevA.101.023822}. Given an initial state of the system described by a density matrix, we aim to find the control field that compels the system to follow a given prescribed dynamics. We denote the ground and excited levels by $\ket{0}$ and $\ket{1}$, respectively. The total Hamiltonian $H(t)$ that describes the system is composed of an unperturbed part $H_0$ and a time-dependent part $H_1(t)$,

\begin{equation}
    \label{h0}
    H(t) = H_0 + H_1(t) = \frac{\omega}{2}(\ket{0}\bra{0} - \ket{1}\bra{1}) - \mu E(t)(\ket{0}\bra{1} + \ket{1}\bra{0}),
\end{equation}
where $\omega$ is the resonance frequency of the two-level system and $\mu$ is the projection of the electric dipole moment along the field polarization axis. The field $E(t)$ is assumed to be a linearly-polarized pulse described by

\begin{equation}
    \label{eq:campo}
    E(t) = \varepsilon(t) e^{-i\omega_p t} + \varepsilon^*(t) e^{i\omega_p t},
\end{equation}
where $\varepsilon(t)$ is a complex function that defines the pulse amplitude and shape, while $\omega_p$ is the carrier frequency of the pulse, with the asterisk denoting the complex conjugate.

Considering the interaction picture and performing the rotating wave approximation, the system turns to be described by the transformed Hamiltonian $\tilde{H}(t)$

\begin{equation}
    \label{Vint}
    \tilde{H}(t) = - \mu (\varepsilon^* (t) e^{i( \omega_p - \omega)t}\ket{0}\bra{1} + \varepsilon(t) e^{-i( \omega_p - \omega)t}\ket{1}\bra{0}).
\end{equation}
We assume that the associated dynamics of the system is given by the Markovian master equation,
\begin{equation}
    \label{eq:lindblad}
    i\dot{\rho}(t) = [\tilde{H}(t), \rho(t)] + \frac{\gamma}{2} D_{deph}[\rho(t)] + \Gamma D_{therm}[\rho(t)],
\end{equation}
where the dot stands for the time derivative and the system is subject to both dephasing and thermal noise with rates $\gamma$ and $\Gamma$, respectively. Atomic units are adopted throughout the work, thus we set $\hbar=1$. The system-environment coupling terms for dephasing $D_{deph}[\rho]$ and thermal noise $D_{therm}[\rho]$ are explicitly given by,  

\begin{align}
      &  D_{deph}[\rho] =  -2(\rho_{01}\ket{0}\bra{1} + \rho_{10}\ket{1}\bra{0}), \\
       &  D_{therm}[\rho] =  -2\left[(\bar{n}\rho_{00}-(\bar{n}+1)\rho_{11}\right]\left(\ket{0}\bra{0} - \ket{1}\bra{1}\right) 
     -(2\bar{n}+1)\left(\rho_{01} \ket{0}\bra{1} +\rho_{10} \ket{1}\bra{0}\right),
\end{align}
where $\bar{n}$ is the average number thermal phonons and we have denoted the elements of the density matrix as $\rho_{ij}$, $i,j = 0,1$.

From the master equation (\ref{eq:lindblad}) and by the fact that $\rho_{00}+\rho_{11}=1$ and $\rho_{01}=\rho_{10}^*$ it follows that the system dynamics is dictated by two coupled differential equations,

\begin{align}
\label{eq:rho00}
  &  \dot{\rho}_{00} = 2 \mu \rm Im\left\{\rho_{01} \varepsilon(t) e^{-i (\omega_p - \omega)t}\right\} + 2\Gamma[\bar{n}+1 +(2\bar{n}-1)\rho_{00}], \\
    \label{eq:rho10}
 &   \dot{\rho}_{01} = -i \mu (2 \rho_{00} -1) \varepsilon^*(t) e^{i (\omega_p - \omega)t}  - \tilde\Gamma \rho_{01},
\end{align}
where $\rm Im \{\cdot\}$ stands for the imaginary part and $\tilde\Gamma = \gamma + \Gamma(2\Bar{n}+1)$.

Inverting Eq.~(\ref{eq:rho10}) in terms of $\varepsilon^*(t)$ and substituting into Eq.~(\ref{eq:campo}), we obtain an expression for the field in terms of the elements of the density matrix,

\begin{equation}
    \label{eq:campo2}
    E(t) = \frac{2 \rm Im[(\dot{\rho}_{10}+\tilde\Gamma \rho_{10})e^{-i\omega t}]}{\mu(2 \rho_{00}-1)}.
\end{equation}
We note that, although up to Eq.(\ref{eq:campo2}) our approach has been very close to Ref.\cite{PhysRevA.100.012103}, from this point forward our results will depart from this reference, since we consider the phase of the coherence to be time-dependent.

Writing the coherence as $\rho_{01} (t) = |\rho_{01} (t)| e^{i\phi(t)}$, substituting in Eqs.(\ref{eq:rho00}) and (\ref{eq:rho10}) and dividing one by the other in such a way to eliminate the dependence on the field, we obtain the following differential equation,

\begin{equation}
    \label{eq: rho01}
    \frac{d}{dt}|\rho_{01} (t)|^2 = (1-2 \rho_{00}(t))\left[\dot{\rho}_{00}(t)+\Gamma_2 \rho_{00}(t) - \Gamma_1\right] - 2\tilde\Gamma |\rho_{01} (t)|^2,
\end{equation}
where $\Gamma_1 \equiv 2 \Gamma(\bar{n} + 1)$ and $\Gamma_2 \equiv 2 \Gamma(2\bar{n} + 1)$.
There is a second relation which follows from the equations of motion (\ref{eq:rho00}) and (\ref{eq:rho10}),

\begin{equation}
\label{eq:phidot}
    |\rho_{01}|\dot{\phi}(t)=\mu(1-2\rho_{00})\rm{Re}\{\varepsilon(t)\rm e^{-i(\omega t-\phi)}\}.
\end{equation}
Note from (\ref{eq:phidot}) that the $|\rho_{01}|\dot{\phi}(t)$ has to vanish whenever $\rho_{00}=1/2$.

 We can now set up a methodology to design the control by noting from Eq.~(\ref{eq:campo2}) that the field can be build up \textit{if} $\rho_{00}(t)$ and $\rho_{01}(t)$ were known. We will instead choose a desired function for the population dynamics $P(t)$ in the place of $\rho_{00}(t)$. We then perform the substitution $\rho_{00}(t)\rightarrow P(t)$ in Eq.~(\ref{eq: rho01}), whose solution will furnish the absolute value of the coherence for the desired population dynamics, which we will denote by $C(t)$, i.e., we write,
 
 \begin{equation}
    \label{eq: C(t)}
    \frac{d}{dt}G(t) = (1-2 P(t))\left[\dot{P}(t)+\Gamma_2 P(t) - \Gamma_1\right] - 2\tilde\Gamma G(t),
\end{equation}
where we have made the identification $ |\rho_{01} (t)|\rightarrow C(t)=\sqrt{G(t)}$.

Finally, we choose a second desired function for the phase of the coherence, replacing $\phi(t)\rightarrow\Phi(t)$. It should be notice that $P(t)$ and $\Phi(t)$ are user-designed function not to be confused with $\rho_{00}$ and $\phi(t)$ which refer to the solution of the equation of motion for a given external field. Also, $|\rho_{01}(t)|$ refers to Eq. (\ref{eq: rho01}), while $C(t)$ refers to Eq.~(\ref{eq: C(t)}).
 
Thus, prescribing a function for the population dynamics $P(t)$, solving Eq.~(\ref{eq: C(t)}) to obtain $C(t)$ and with an additional prescribed function for the coherence phase $\Phi(t)$, the equation for the field (\ref{eq:campo2}) can be expressed in terms of these three dynamics functions with the substitutions $\rho_{00}(t)\rightarrow P(t)$, $\phi(t)\rightarrow\Phi(t)$ and $|\rho_{01}(t)|\rightarrow C(t)$. After some algebra, the field can be expressed in a convenient sinusoidal form,

\begin{equation}
    \label{eq:campodiss}
    E(t) = A(t) \sin(\omega t + \Phi(t) + \Lambda(t)),
\end{equation}
where the time-dependent phase $\Lambda(t)$ is given by,
\begin{equation}
    \label{eq:lambdadiss}
    \Lambda(t) = \tan^{-1} \left[\frac{\dot{\Phi}(t)C(t)}{\dot{C}(t) + \tilde\Gamma C(t)}\right],
\end{equation}
and the time-dependent amplitude is given by,
\begin{equation}
    \label{eq:A(t)}
    A(t) = \frac{2}{\mu} \xi(t) \sqrt{\frac{(\dot{P}(t)+\Gamma_1 P(t) - \Gamma_2)^2}{4C^2(t)} + \frac{\dot{\Phi}^2(t)C^2(t)}{(2P(t)-1)^2}},
\end{equation}
with $\xi(t) = {\rm sgn}\left(\frac{\dot{P}(t)+\Gamma_1 P(t) - \Gamma_2}{2C(t)}\right) $.

We have the additional constraint for choosing $\Phi(t)$ resulting from Eq. (\ref{eq:phidot}),

\begin{equation}\label{eq:phasecond}
    \dot{\Phi}(t^*)=0, \quad \text{for} \quad P(t^*)=1/2.
\end{equation}
Strictly, one should have $C(t^*)\dot{\Phi}(t^*)=0$ but we can take advantage of the flexibility of choosing $\Phi(t)$. Thus, we set the derivative of the phase of the coherence to vanish whenever the populations are equal. Note that with this choice, there is no singularity in the second term in the square root of Eq. (\ref{eq:A(t)}). A similar condition has also been found in the case of the unitary dynamics of pure states \cite{PhysRevA.110.022201}.

\section{The absolute value of the coherence}

We consider the solution of Eq.~(\ref{eq: C(t)}), which furnishes the absolute value of the coherence $C(t)$ once the population dynamics $P(t)$ is defined. Introducing the \textit{ansatz} $G(t)=\rm{e}^{-2\tilde\Gamma t}F(t)$, we obtain the following differential equation for $F(t)$,
\begin{equation}
    \label{eq: edoF(t)}
    \frac{d}{dt}F(t) = \rm{e}^{2\gamma t}\frac{d}{dt}\left[P(t)-P^2(t)\right]-\rm e^{2\tilde\Gamma t}\left[2\Gamma_2 P(t)^2-(2\Gamma_1+\Gamma_2)P(t)+\Gamma_1\right],
\end{equation}
which can be integrated yielding
\begin{equation}
    \label{eq: F(t)}
    F(t) = \rm{e}^{2\tilde\Gamma t}\left(P(t)-P(t)^2\right)-k-\tilde\gamma\left(\rm e^{2\tilde\Gamma t}-1\right)+\tilde\Gamma_1I_1(\tilde\Gamma,t)+\tilde\Gamma_2I_2(\tilde\Gamma,t)
\end{equation}
where we have defined the constants $k=P(0)-P(0)^2-C(0)^2$, $\tilde\gamma=\Gamma_1/(2\tilde\Gamma)$, $\tilde\Gamma_1=2\Gamma_1+\Gamma_2-2\tilde\Gamma$ and $\tilde\Gamma_2=2(\tilde\Gamma-\Gamma_2)$, while we have set the integrals $I_1(\gamma,t)$ and $I_2(\gamma,t)$ as
\begin{equation}
\label{eq:I1}
    I_1(\tilde\Gamma,t)=\int_0^{t}P(t')\rm{e}^{2\tilde\Gamma t'}dt'
\end{equation}
and
\begin{equation}
\label{eq:I2}
    I_2(\tilde\Gamma,t)=\int_0^{t}P(t')^2\rm{e}^{2\tilde\Gamma t'}dt'.
\end{equation}  

Thus, we can express the absolute value of the coherence as,
\begin{equation}
    \label{eq: G(t)}
  C(t)^2 = \left(P(t)-P(t)^2\right) + \rm e^{-2\tilde\Gamma t} \left(\tilde\gamma - k\right) - \tilde\gamma + e^{-2\tilde\Gamma t}\left[\tilde\Gamma_1I_1(\tilde\Gamma,t)+\tilde\Gamma_2I_2(\tilde\Gamma,t)\right].
\end{equation}
Note that the prescribed population dynamics $P(t)$ has to be chosen to meet the condition $C(t)\ge0$, for all $t$, which limits the possible transitions. The populations that can be reached from a given initial one are inferred from the right-hand-side of Eq.~(\ref{eq: G(t)}) which must not be negative.

We observe that the constant $k$ is determined from the initials population and coherence, as defined above. Additionally, the initial coherence $C(0)$ is not completely arbitrary, since it has to satisfy the relation $C(0)^2\leq (P(0) - P(0)^2)$ to guarantee that ${\rm Tr}\rho^2\leq1$, this fact implies that $0\leq k\leq1/4$. Note that for $k=0$ we have an initially pure state and otherwise a mixed state. In fact, $k$ is related to the initial purity by ${\rm{Tr}}\{\rho(0)^2\}=1-2k$.

\subsection*{Constant population}

By specifying the function $P(t)$, the integrals (\ref{eq:I1}) and (\ref{eq:I2}) may be calculated analytically, and the control field can be expressed by an analytical formula. For instance, consider the situation when it is desired to keep the population constant, i.e., $P(t)=P$. In this case, the integrals in Eq.(\ref{eq: G(t)}) are easily calculated and we can write the coherence as,

\begin{equation}
    C^2(t)=C_{\infty}^2+(\lambda(P)-k){\rm e}^{-2\tilde{\Gamma}t},
\end{equation}
where $\lambda(P)= \tilde{\gamma}-P(\tilde{\Gamma}_1+P\tilde{\Gamma}_2)/2\tilde{\Gamma}$ and the asymptotic value of the coherence $C_{\infty}$ is expressed as,

\begin{equation}\label{eq:cinfty}
  C_{\infty}^2=P-P^2-\lambda(P).
\end{equation}
Thus, if the population is kept constant, the initial coherence $C(0)$ decays exponentially to the asymptotic value $C_{\infty}$ in a typical timescale of $1/(2\tilde{\Gamma})$. Additionally, if initial coherence and population are set such that $k=\lambda$, then the absolute value of the coherence will remain constant. It is then possible to produce steady states even in the presence of dephasing and thermal noise using the appropriate field given by Eq.(\ref{eq:campodiss}). The present analysis is independent of the choice of the phase of the coherence $\Phi(t)$, which may or may not vary. Note also that if only dephasing is present, i.e., $\Gamma=0$ then $C_{\infty}=0$ and the coherence always vanish for sufficiently long time

\subsection*{Unitary evolution}

It is also instructive to consider the particular case of unitary evolution, i.e., $\Gamma=\gamma=0$. In this case, we have simply

\begin{equation}
    \label{eq:h(t)analitica}
    C^2(t) = P(t) - P(t)^2 - k.
\end{equation}
Thus, the prescribed population dynamics $P(t)$ for the unitary case has to be chosen to meet the condition $P(t)-P(t)^2\ge k$ for all $t$, i.e.,

\begin{equation}
  \label{condicao_P}
  1/2-\sqrt{1/4-k}\leq P(t)\leq 1/2+\sqrt{1/4-k}  .
\end{equation}
Finally, Eq.(\ref{eq:h(t)analitica}) can be used to control the final absolute value of the coherence through the appropriate choice of the final value $P(t_f)$.


\section{Simultaneous control of relative phase and populations}

\begin{figure}[ht]
\includegraphics[width=10cm]{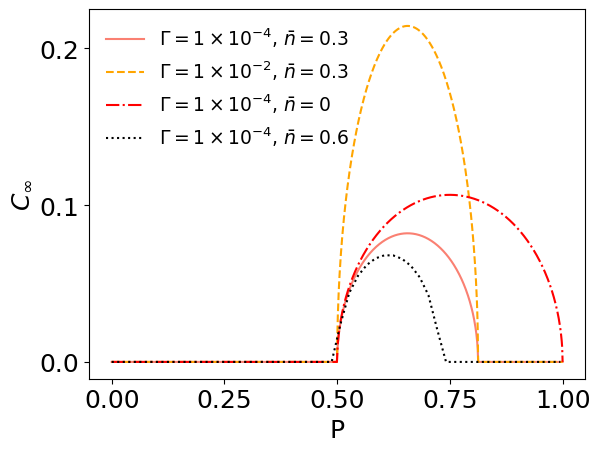}
\centering
 \caption{Asymptotic value of the coherence $C_{\infty}$ as a function of the population. The rate of dephasing is fixed in $\gamma=0.001$. The real part of $C_{\infty}$ is shown for several values of $\Gamma$ and photon numbers as given by Eq.~(\ref{eq:cinfty}). }\label{fig:comparação_nbarra}
\end{figure}

Throughout this work, we perform simulations using $\omega=0.02$ and $\mu = 6$, which corresponds to a charge migration problem and allows for comparisons with previous works\cite{PhysRevA.100.012103,PhysRevA.101.023822,10.1063/1.2970088}. We start illustrating our results considering constant population $P(t)=P$. Figure~\ref{fig:comparação_nbarra} shows the real part of $C_{\infty}$ calculated by means of Eq.~(\ref{eq:cinfty}) as a function of the population for several values of the thermal noise rate $\Gamma$ and photon number $\bar{n}$ and fixed dephasing rate $\gamma=0.001$. The positive values of $C_{\infty}$ indicate the populations that can be maintained constant indefinitely by the control field. We note that the peak value of the asymptotic coherence depends essentially of the magnitude of $\Gamma$, while the range of possible values of $P$ depends of $\bar{n}$. In fact, solving  $C_{\infty}=0$ for $P$, we find two roots, $P=1/2$ and $P=(\bar{n}+1)/(2\bar{n}+1)$, so that $P=1$ can be kept constant only for zero temperature, $\bar{n}=0$.

\begin{figure}[ht]
\includegraphics[width=15cm]{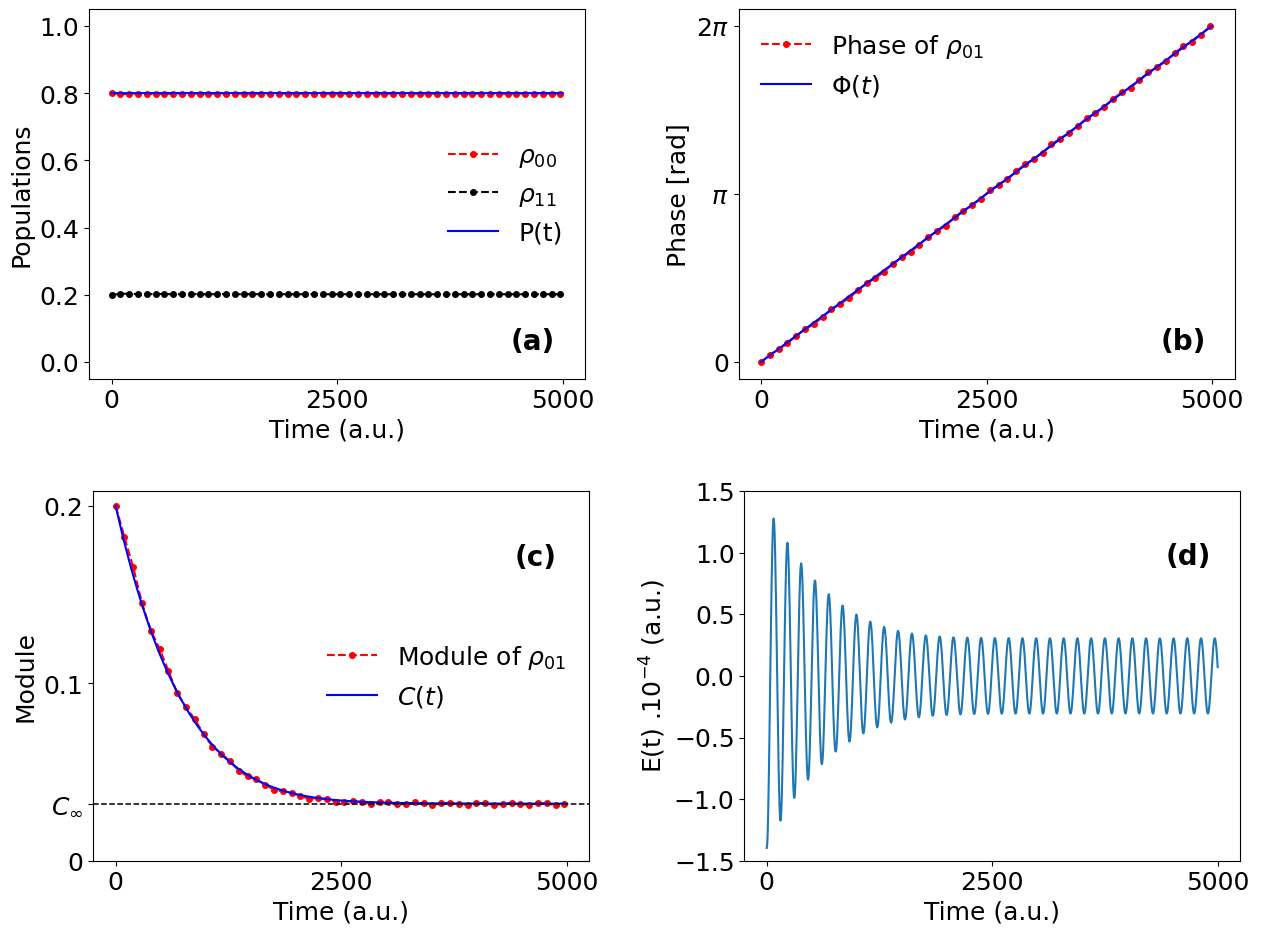}
\centering
 \caption{Comparison between the chosen functions for population and coherence phase with the exact numerical calculations. Case of constant population $P(t)=0.8$ and linear phase $\Phi(t)=\alpha t$ with $C(0)=0.2$, $t_f=5000$ and $\Phi(t_f)=2\pi$ . (a) Population dynamics; (b) Coherence phase; (c) Absolute value of the coherence (the dotted line indicates the value of $C_{\infty}$); (d) Control field obtained from Eq.~(\ref{eq:campodiss}).}\label{fig:fase_var}
\end{figure}

Figure~\ref{fig:fase_var} illustrates the case for $P=0.8$, $C(0)=0.2$ and $t_f=5000$. We choose a linear function for the coherence phase $\Phi(t)=\alpha t$, such that the final phase corresponds to a complete rotation in the Bloch sphere, $\Phi(t_f)=2\pi$. Fig.~\ref{fig:fase_var}(a) compares the intended constant population $P=0.8$ with full numerical calculation, which is obtained evolving the system with the designed control field out of the RWA. Fig.~\ref{fig:fase_var}(b) compares also the numerically obtained phase with the chosen linear function for $\Phi(t)$. Fig.~\ref{fig:fase_var}(c) shows the decaying of the coherence modulus to its asymptotic value $C_{\infty}$, indicated by the dashed line. Finally in Fig.~\ref{fig:fase_var}(d) the control field is shown. Note that after a transient the field reaches a constant amplitude and constant frequency $\omega+\alpha$.

As a second illustrative example, now involving changing of the population, we choose a sine squared function for $P(t)$,
\begin{equation}
    \label{eq:P(t)_deph}
    P(t) =  \left(P_f - P_i \right) \sin^2\left(\frac{\pi t}{2t_f}\right) + P_i
\end{equation}
where $P_i$ e $P_f$ stands for the initial and final populations of the ground level, respectively, with $0\le t\le t_f$. This choice is convenient since it allows to calculate $C(t)$ in Eq.~(\ref{eq: G(t)}) analytically. The obtained formula is given in the appendix.


Figure \ref{fig:mapa_dephasing} depicts the accessible regions for transitions when thermal noise is absent ($\Gamma=0$) for different values of $\gamma$. Each point on the map corresponds to a transition from an initial population $P_i$ to a final population $P_f$ of the ground state. The white regions indicate the inaccessible region where the RHS of Eq.~(\ref{eq:h(t)analitica}) yields negative values, while the light-gray regions indicate the accessible region for the unitary case ($\gamma=\Gamma=0$). We emphasize that the overall hourglass shape is due to the fact that each panel is drawn for a fixed value of the initial coherence $C(0)$, consequently with distinct purities, see Eq.~(\ref{eq:h(t)analitica}). The dark-gray regions indicate the accessible for $\gamma\neq0$, where the RHS of Eq.~(\ref{eq: G(t)}) yields non-negative values. In the present case, these dark-gray regions are always contained within the light-gray regions. In the first column of Fig. \ref{fig:mapa_dephasing}, panels (a), (b) and (c), $C(0)=0.02$ and $t_f=1500$. In the second column, panels (d), (e) and (f) , $C(0)=0.2$ and $t_f=1500$. In the third column, panels (g), (h) and (i),  $C(0)=0.2$ and $t_f=3500$. In the first row of the figure, panels (a), (d) and (g), $\gamma=1\times10^{-4}$. In the second row, panels (b), (e) and (h), $\gamma=1\times10^{-3}$. In the third row, panels (c), (f) and (i), $\gamma=5\times10^{-3}$. Note, from panels (a), (b) and (c), that the transition from $P_i=0.9$ to $P_f=0.4$ is allowed for $\gamma\le 10^{-3}$ but it is not allowed for $\gamma=5\times10^{-3}$. Comparing the rows, one can note how the accessible regions turns smaller as the dephasing rate increases. We can also observe the effects on the accessible regions of the initial value of the coherence contrasting the panels on the first and on the second columns of the figure: the hourglass shape gains waist, while shrinks at the top and the bottom values of $P_i$ as the initial coherence increases. Finally, the result of increasing the final time is perceived comparing the two last columns: in general, increasing the time is detrimental for the accessible region, since the dephasing has more time to act on the system. 

\begin{figure}[h!]
\includegraphics[width=14cm]{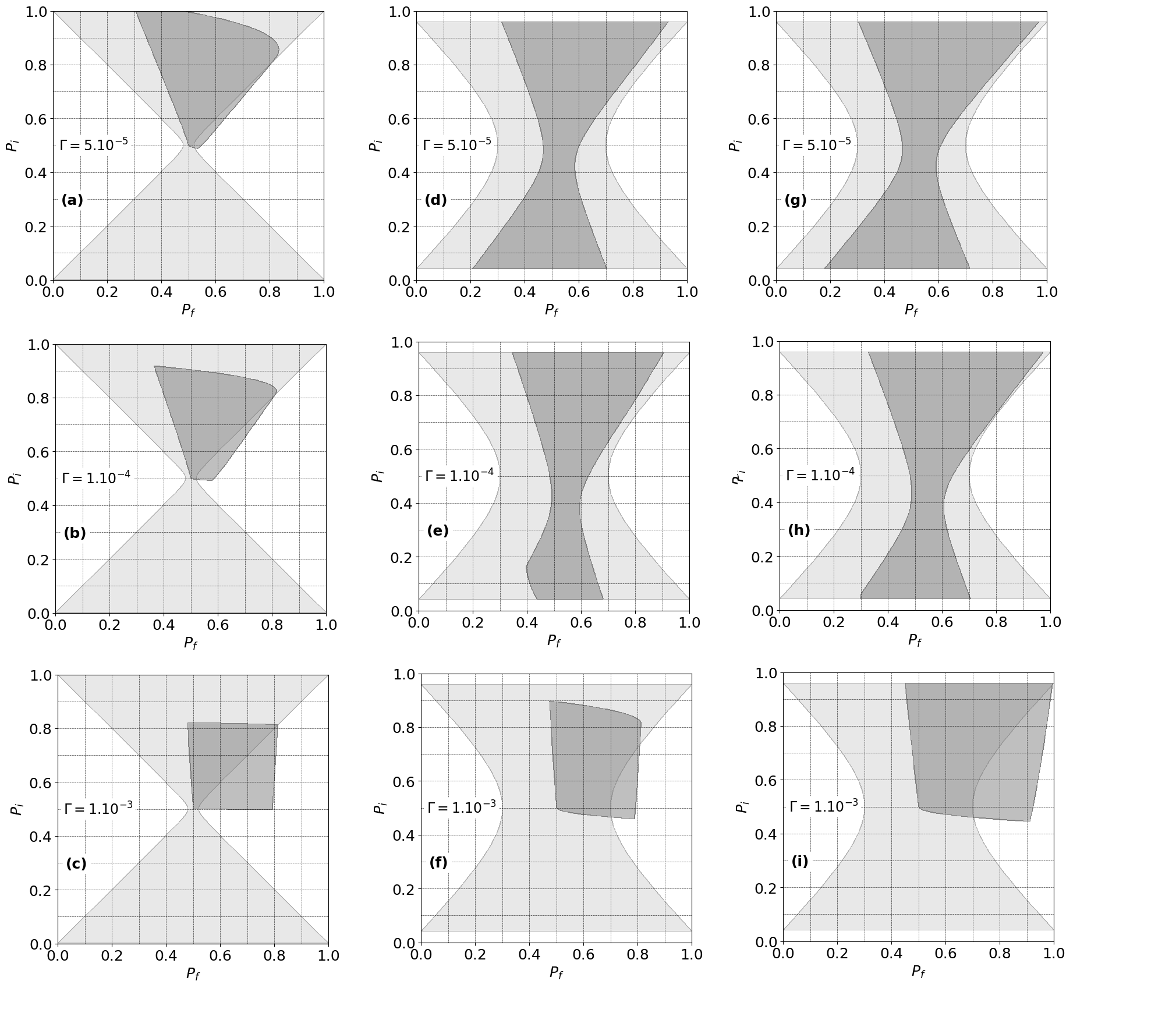}
\centering
 \caption{Comparison among the accessible regions for transition between the initial population $P_i$ to the final population $P_f$ for different values of the thermal noise rate $\Gamma$. $P(t)$ chosen as a sine squared function, Eq.~(\ref{eq:P(t)_deph}) White regions are inaccessible and light-gray are accessible regions for $\gamma=\Gamma=0$. Dark-gray accessible region for $\Gamma\neq0$ and fixed dephasing rate $\gamma=10^{-3}$ and $t_f = 1500$.(a) $\Gamma=5\times10^{-5}$, $\bar{n} = 0.3$ and $C(0) = 0.02$, (b) $\Gamma=10^{-4}$, $\bar{n} = 0.3$ and $C(0) = 0.02$, (c) $\Gamma=10^{-3}$, $\bar{n} = 0.3$ and $C(0) = 0.02$.  (d) $5\times\Gamma=10^{-5}$, $\bar{n} = 0.3$ and $C(0) = 0.2$, (e) $\Gamma=10^{-4}$, $\bar{n} = 0.3$ and $C(0) = 0.2$, (f) $\Gamma=10^{-3}$, $\bar{n} = 0.3$ and $C(0) = 0.2$; (g) $5\times\Gamma=10^{-5}$, $\bar{n} = 0$ and $C(0) = 0.2$, (h) $\Gamma=10^{-4}$, $\bar{n} = 0$ and $C(0) = 0.2$, (i) $\Gamma=10^{-3}$, $\bar{n} = 0$ and $C(0) = 0.2$.}\label{fig:mapa_therml}
\end{figure}

Figure \ref{fig:mapa_therml} shows the accessible transitions when both dephasing and thermal noise are present  for some values of $\Gamma$ and $\bar{n}$. As in Fig.~\ref{fig:mapa_dephasing}, each point on the map connects an initial population $P_i$ to a final population $P_f$, with the white regions indicating the inaccessible region and the light-gray regions indicating the accessible region for the unitary case. The dark-gray regions indicate the accessible for $\gamma=10^{-3}$ and $\Gamma\neq0$, where the RHS of Eq.~(\ref{eq: G(t)}) yields non-negative values. The final times and initial coherences are set with the same values of Fig.~\ref{fig:mapa_dephasing} in the corresponding panels. Note that in this case the dark-gray region is not always contained within the light-gray region. In the first row of the figure, panels (a), (d) and (g), $\Gamma=5\times10^{-5}$. In the second row, panels (b), (e) and (h), $\Gamma=1\times10^{-4}$. In the third row, panels (c), (f) and (i), $\Gamma=1\times10^{-3}$. From panel (a) to (f) the mean number of thermal photons is $\bar{n}=0.3$, while $\bar{n}=0$ in panels (g), (h) and (i).  Comparing the rows, one can note how the accessible regions shrink and move as the thermal noise rate increases. The propensity to allow transitions between high values of ground-state population can be justified by the fact that the thermal noise generally tends to drive the system to the ground level. Contrasting the last two columns, we can observe how decreasing the mean number of thermal photons leads to the growth of the accessible region.

\begin{figure}[ht!]
\includegraphics[width=8cm]{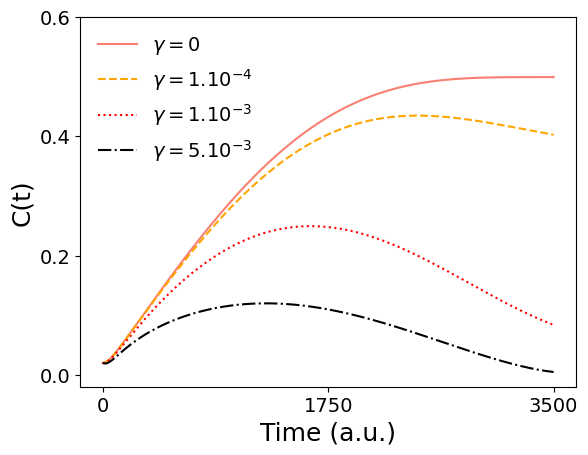}
\centering
 \caption{Absolute value of the coherence $C(t)$ for different values of $\gamma$ and no thermal noise $\Gamma=0$ calculated from Eq.~(\ref{eq: G(t)}) with $P(t)$ given by (\ref{eq:P(t)_deph}) and parameters $P_i = 1$, $P_f = 0.5001$,  $t_f = 3500$ e $C(0) = 0.02$.}\label{fig:C_dephasing}
\end{figure}

\begin{figure}[hb!]
\includegraphics[width=8cm]{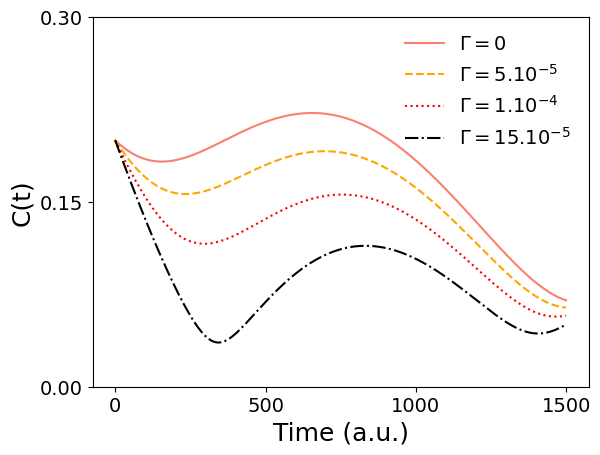}
\centering
 \caption{The same of Fig.~\ref{fig:C_dephasing} for different values of $\Gamma$ with $P_i = 0.2$, $P_f = 0.6$,  $t_f = 1500$, $C(0) = 0.2$ and $\gamma = 0.001$.}\label{fig:C_thermal}
\end{figure}

In Figs.~\ref{fig:C_dephasing} and \ref{fig:C_thermal}, we show how the absolute values of the coherence is affected by the dephasing and the thermal noise rates, respectively. We observe an overall decreasing of the coherence for increasing rates. However, this reduction in the coherence is not uniform over time, indicating that the final time can be adjusted to maintain some coherence in the final state.



\pagebreak

\begin{figure}[h!]
\includegraphics[width=14cm]{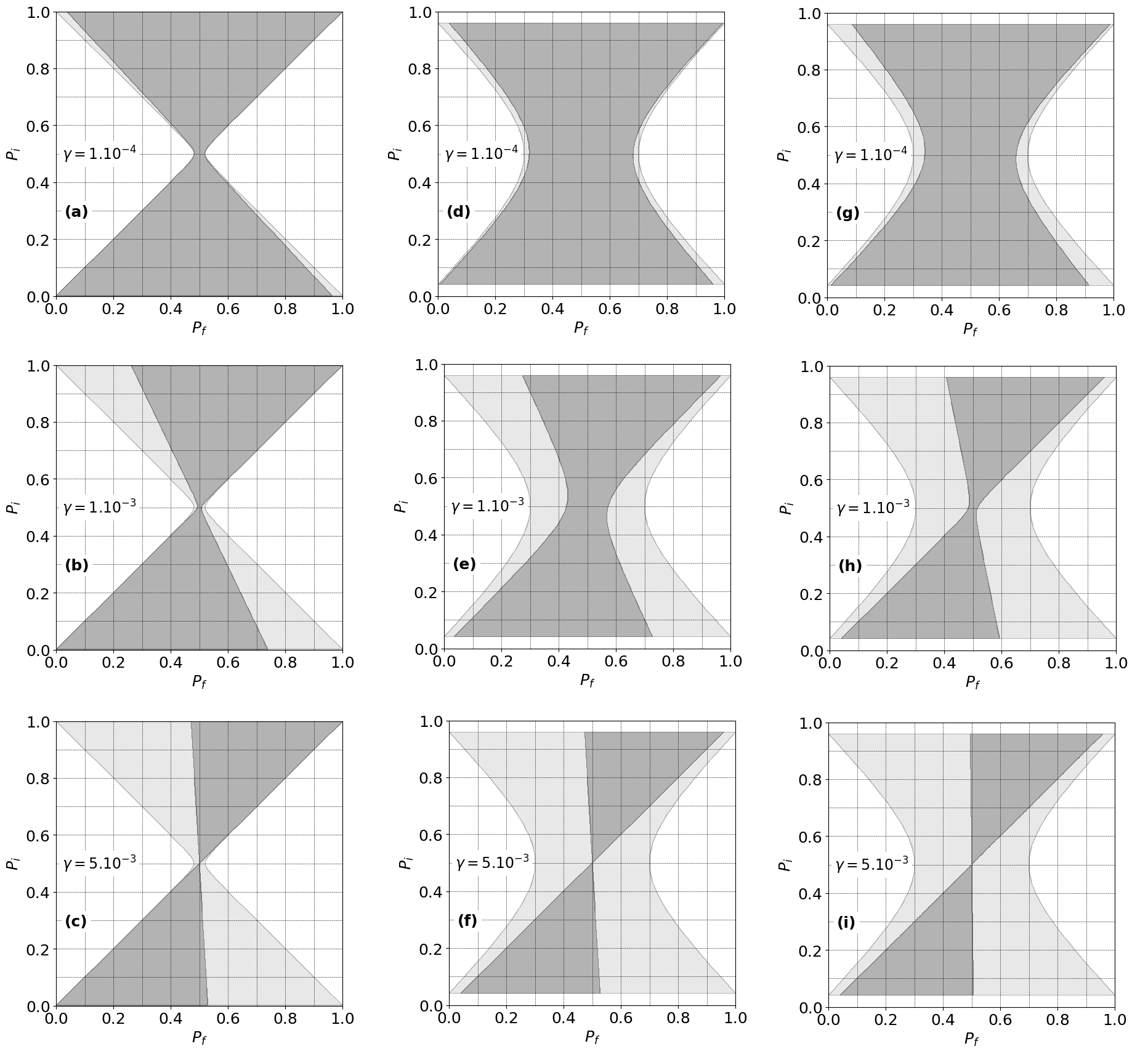}
\centering
 \caption{Comparison among the accessible regions for transition between the initial population $P_i$ to the final population $P_f$ for different values of $\gamma$ and no thermal noise $\Gamma=0$. $P(t)$ chosen as a sine squared function, Eq.~(\ref{eq:P(t)_deph}). White regions: inaccessible and light-gray accessible regions for $\gamma=0$. Dark-gray accessible region for $\gamma\neq0$. (a) $\gamma=10^{-4}$, $t_f = 1500$ and $C(0) = 0.02$, (b) $\gamma=10^{-3}$, $t_f = 1500$ and $C(0) = 0.02$, (c) $\gamma=5\times10^{-3}$, $t_f = 1500$ and $C(0) = 0.02$.  (d) $\gamma=10^{-4}$, $t_f = 1500$ and $C(0) = 0.2$, (e) $\gamma=10^{-3}$, $t_f = 1500$ and $C(0) = 0.2$, (f) $\gamma=5\times10^{-3}$, $t_f = 1500$ and $C(0) = 0.2$; (g) $\gamma=10^{-4}$, $t_f = 3500$ and $C(0) = 0.2$, (h) $\gamma=10^{-3}$, $t_f = 3500$ and $C(0) = 0.2$, (i) $\gamma=5\times10^{-3}$, $t_f = 3500$ and $C(0) = 0.2$.}\label{fig:mapa_dephasing}
\end{figure}

\pagebreak

\pagebreak


\pagebreak

\begin{figure}[ht]
\includegraphics[width=15cm]{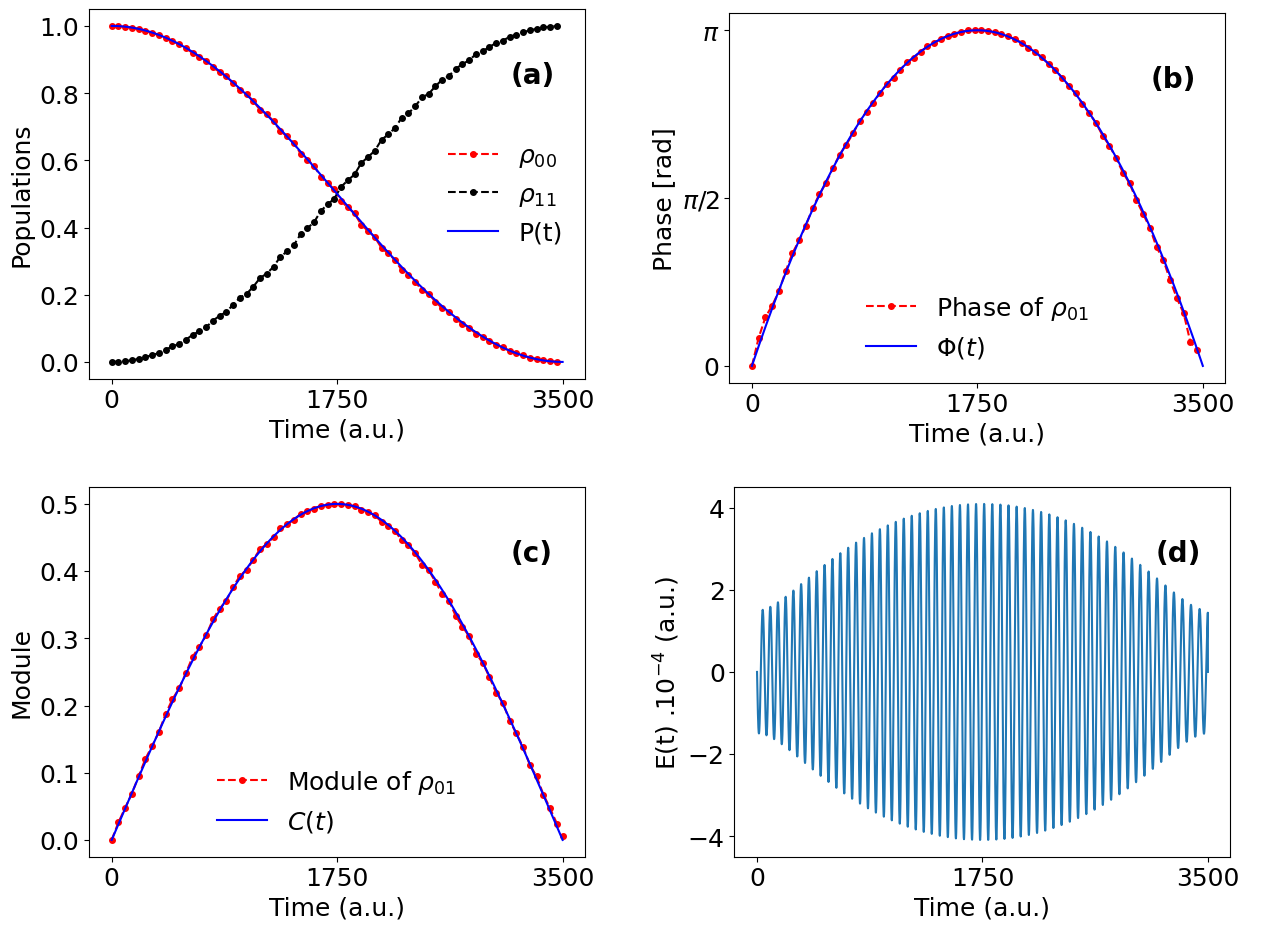}
\centering
 \caption{(a) Comparison between the population control function $P(t)$ given in (\ref{eq:P(t)_deph}) and the numerical simulation of populations; (b) comparison between the phase control function $\Phi(t)$ given by a quadratic function and the numerically calculated phase; (c) the function $C(t)$ calculated from (\ref{eq: G(t)}) compared to the numerically calculated $|\rho_{01}|$; (d) the control field calculated from the expression (\ref{eq:campodiss}). Parameters used: $\gamma =\Gamma =\bar{n} = 0, C(0) = 0$. The initial and final ground state population are set to $P_i = 1$ and $P_f = 0$, respectively and the initial and final phases to $\Phi(0) =\Phi(t_f) = 0$. The final time is $t_f = 3500$.}\label{fig:controle_unitario}
\end{figure}

We now illustrate the simultaneous control of population and coherence phase for the unitary case by setting, along with $P(t)$, the function $\Phi(t)$. We consider $\Phi(t)$ as a parabola with a maximum at the time when the populations of the levels are equal. Figure \ref{fig:controle_unitario} illustrates this case where we set the initial and final populations as $P_i=1$ and $P_f=0$ and the initial and final phases as $\Phi(0)=0$ and $\Phi(t_f)=0$, with $\gamma=\Gamma=0$, which means a population inversion for the unitary dynamics. The chosen functions $P(t)$ and $\Phi(t)$ are substituted in Eq.~(\ref{eq:campodiss}), with $C(t)$ calculated from (\ref{eq:h(t)analitica}). The choice of $P_i=1$ implies that $C(0)=0$, while the choice of $P_f=0$ and the final time as $t_f=3500$, leads to $t^*=t_f/2=1750$, see Eq.~(\ref{eq:phasecond}). Fig.~\ref{fig:controle_unitario}(a) shows the population dynamics, where we compare the function $P(t)$ with the dynamics of $\rho_{00}$ obtained from the direct numerical solution of the master equation (in this case, the Heisenberg equation) without the RWA. We can note a very good agreement apart some oscillation due to the fact that the engineered field took into account the RWA. The corresponding comparison of the numerically calculated coherence phase and $\Phi(t)$ is given in Fig.~\ref{fig:controle_unitario}(b), which again evidences the very good accordance. Fig.~\ref{fig:controle_unitario}(c) shows the absolute value of the coherence, which has a maximum when the populations are equal, i.e., at $t=t^*$ and returns close to its initial value at $t=t_f$. Finally, Fig.~\ref{fig:controle_unitario}(d) shows the obtained chirped control pulse with symmetric shape relative to $t=t^*$ caused by the choice of the $\Phi(t)$. Note that the shape of the pulse envelope is dictated by the choice of the $\dot{P}(t)$ and $\dot{\Phi}(t)$ and that although $\dot{P}(t)$ is zero at $t=0$ and $t=t_f$, $\dot{\Phi}(t)$ does not approach zero as  $t\rightarrow0$ and $t\rightarrow t_f$, see Eq.~(\ref{eq:A(t)}).

.



\pagebreak

\pagebreak

\begin{figure}[ht]
\includegraphics[width=15cm]{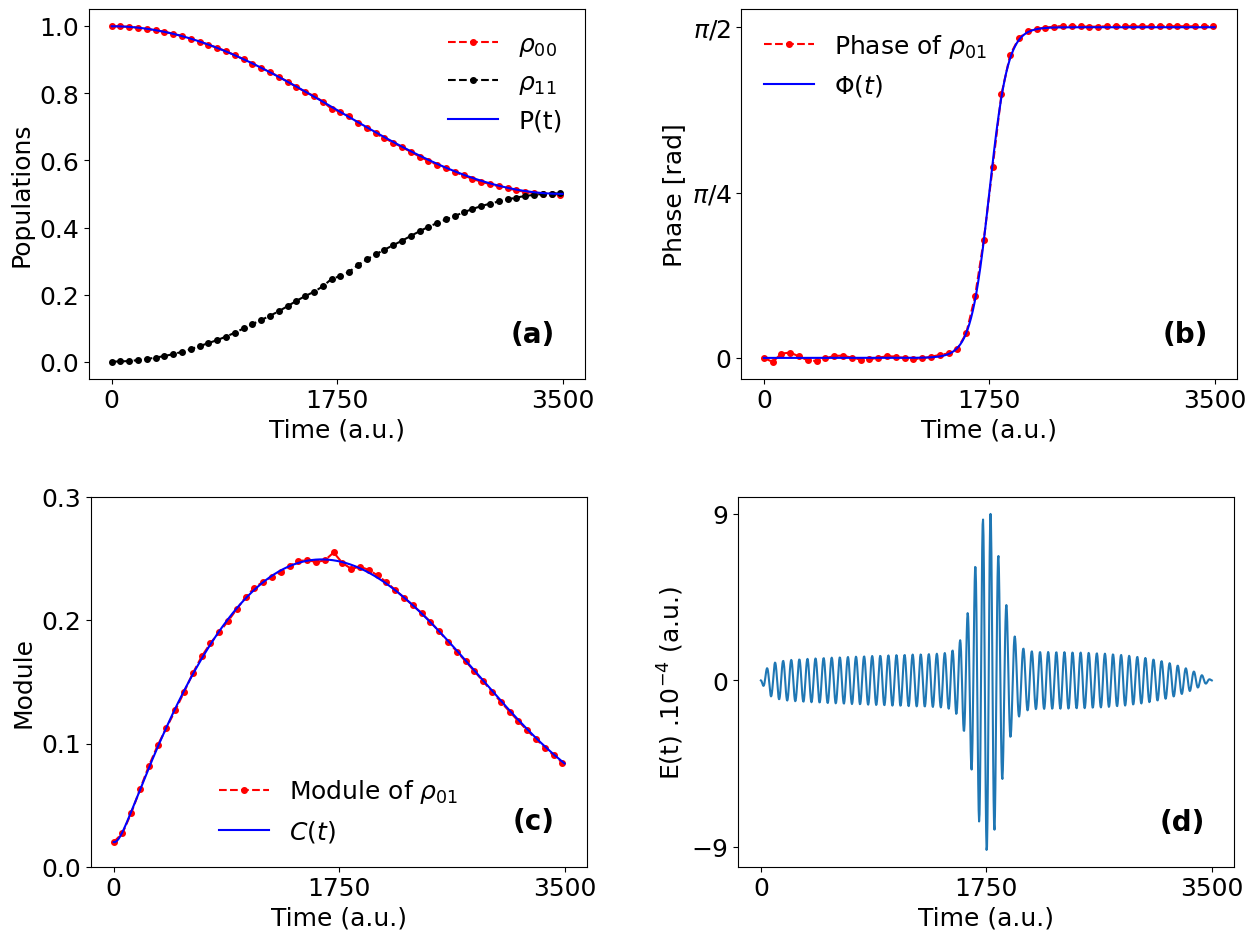}
\centering
 \caption{Comparison between the population control function $P(t)$ given in (\ref{eq:P(t)_deph}) and the numerical simulation of populations (a), comparison between the phase control function $\Phi(t)$ given in (\ref{eq:phi_deph}) and the numerical simulation of the phase (b), the function $C(t)$ calculated from (\ref{eq: G(t)}) (c), and the control field calculated from the expression (\ref{eq:campodiss}) (d). Parameters used: $\gamma = 0.001$, $\Gamma = 0$, $\bar{n} = 0$, $\mu = 6$, $\omega = 0.1$, $C(0) = 0$. The population $P(t)$ transitions from $P_i = 1$ to $P_f = 0.5$. The phase $\Phi(t)$ transitions from $\Phi_i = 0$ to $\Phi_f = \pi$. The simulation was performed for a total time of $t_f = 3500$.}\label{fig:controle_dephasing}
\end{figure}

Figure \ref{fig:controle_dephasing} shows the control of population and coherence phase in the presence of dephasing but with no thermal noise. We set the initial and final population as $P_i=1$, $P_f=0.5$. Note that the choice of $P_f=0.5$ implies that $t^*=t_{f}$, while the choice of $P_i=1$ means that the initial coherence is $C(0)=0$. Additionally, we chose an hyperbolic tangent function for $\Phi(t)$ given by,

\begin{equation}\label{eq:phi_deph}
    \Phi(t)=\frac{\Phi_f-\Phi_i}{2}\tanh(\beta t+\chi)+\frac{\Phi_f+\Phi_i}{2},
\end{equation}
where $\chi=0.5\ln[(1+\sigma)/(1-\sigma)]-\beta t^*$, with $\sigma=(1-\Phi_f-\Phi_i)/(\Phi_f-\Phi_i)$. The initial and final values of the phase are given by the parameters $\Phi_i=\Phi(0)$ and $\Phi_f=\Phi(t_f)$ and set $\Phi_i=0$ and $\Phi_f=\pi/2$, with $\gamma=10^{-3}$ and  $\Gamma=0$. Again, the chosen functions $P(t)$ and $\Phi(t)$ are substituted into Eq.~(\ref{eq:campodiss}), while $C(t)$ is obtained from (\ref{eq: G(t)}). We have set the final time as $t_f=3500$. Fig.~\ref{fig:controle_dephasing}(a) shows the population dynamics, where we compare the function $P(t)$ with the dynamics of $\rho_{00}$ obtained from the direct numerical solution of the Lindblad equation without the RWA. We can note again very good agreement between the chosen population dynamics and the exact dynamics. The corresponding comparison of the numerically calculated coherence phase and $\Phi(t)$ is given in Fig.~\ref{fig:controle_dephasing}(b), which again evidences the very good accordance. Fig.~\ref{fig:controle_dephasing}(c) shows the absolute value of the coherence, which increases at $t=t_f$ with respect to its value at the initial time. As for the unitary case, see Eq.~(\ref{eq:A(t)}), with only dephasing present, $\Gamma_1=\Gamma_2=0$, and the shape of the pulse is again dictated by the derivatives of $P(t)$ and $\Phi(t)$. Note that both approach zero at $t\rightarrow0$ and $t\rightarrow t_f$ defining the overall pulse shape. Fig.~\ref{fig:controle_dephasing}(d) shows the obtained chirped control pulse which has a larger amplitude around $t=t^*$ due to the large value of $\dot{\Phi}(t)$. Also note that a frequency modulation occurs at that time in order to promote the changing of the phase, see Eq.~(\ref{eq:lambdadiss}).

\pagebreak

\begin{figure}[ht]
\includegraphics[width=15cm]{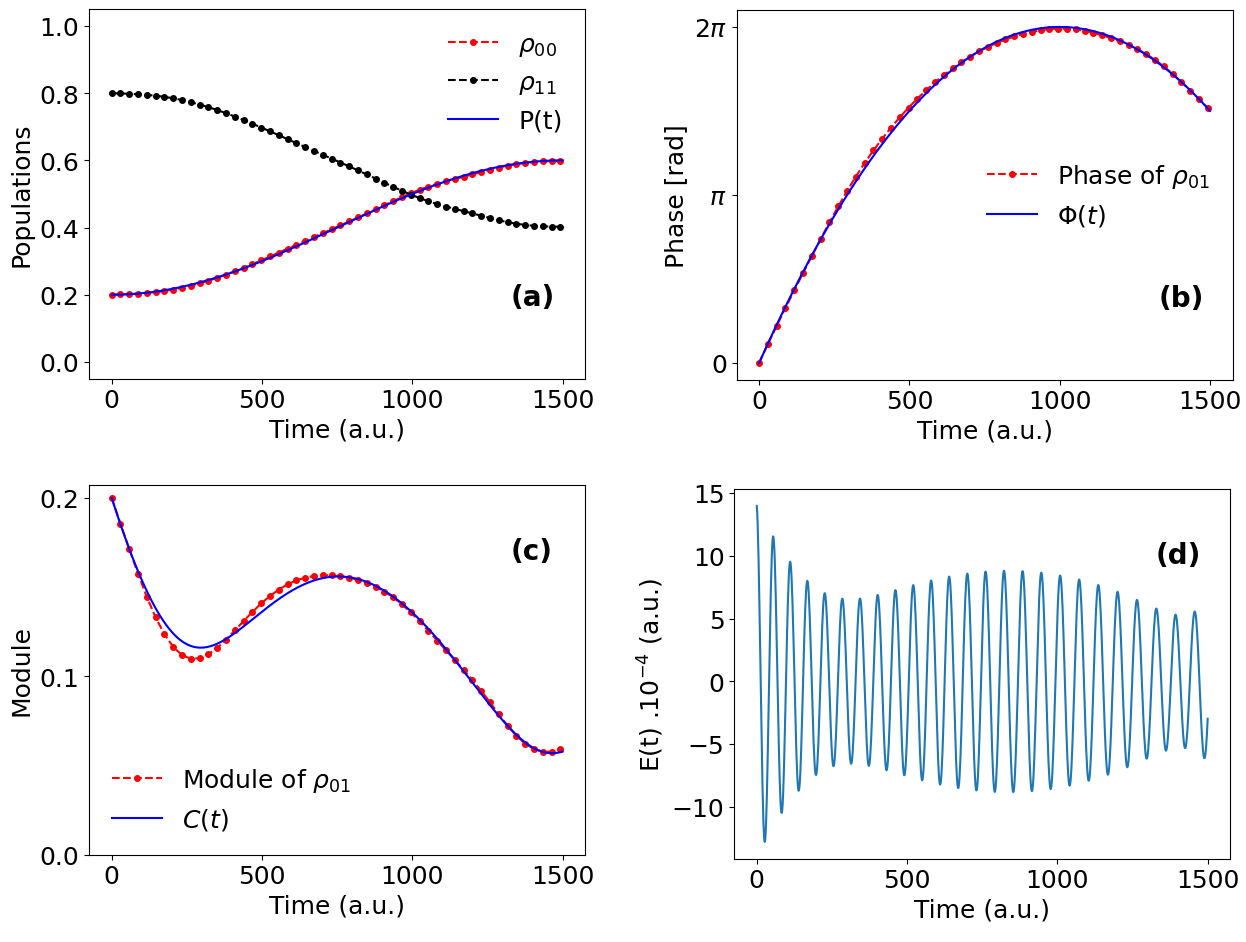}
\centering
 \caption{Comparison between the population control function $P(t)$ given in (\ref{eq:P(t)_deph}) and the numerical simulation of populations (a), comparison between the phase control function $\Phi(t)$ given in (equação quadratica) and the numerical simulation of the phase (b), the function $C(t)$ calculated from (\ref{eq: G(t)}) (c), and the control field calculated from the expression (\ref{eq:campodiss}) (d). Parameters used: $\gamma = 0.001$, $\Gamma = 0.0001$, $\bar{n} = 0.3$, $\mu = 6$, $\omega = 0.1$, $C(0) = 0.2$. The population $P(t)$ transitions from $P_i = 0.2$ to $P_f = 0.6$. The phase $\Phi(t)$ transitions from $\Phi_i = 0 $ to $\Phi_f = \frac{3\pi}{2}$. The simulation was performed for a total time of $t_f = 1500$.}\label{fig:controle_thermal}
\end{figure}

Figure \ref{fig:controle_thermal} presents the control of population and coherence phase in the presence of both dephasing and thermal noise. We choose the initial and final populations as $P_i=0.2$, $P_f=0.6$. The function $\Phi(t)$ is again chosen as a parabola with $\Phi(0)=0$ and $\Phi(t_f)=3\pi/2$. The rates of dephasing and thermal noise are $\gamma=10^{-3}$ and $\Gamma=10^{-4}$. The final time is adjusted to $t_f=1500$, while the initial coherence is set to $C(0)=0.2$. The choice of initial and final populations $P_i$ and $P_f$ implies that $t^*=1000$. The quadratic polynomial form of $\Phi(t)$ is flexible enough to satisfy the initial and final values of the phase having still a free parameter, allowing to adapt the $\Phi(t)$ to meet $\dot{\Phi}(t^*)=0$. Once more, the chosen functions $P(t)$ and $\Phi(t)$ are substituted into Eq.~(\ref{eq:campodiss}), while $C(t)$ is obtained from (\ref{eq: G(t)}). Fig.~\ref{fig:controle_thermal}(a) shows the population dynamics, where we compare the function $P(t)$ with the dynamics of $\rho_{00}$ obtained from the direct numerical solution of the Lindblad equation without the RWA. We can note again very good agreement between the chosen population dynamics and the exact dynamics. The corresponding comparison of the numerically calculated coherence phase and $\Phi(t)$ is given in Fig.~\ref{fig:controle_dephasing}(b), which again evidences the very good accordance. Fig.~\ref{fig:controle_thermal}(c) shows the absolute value of the coherence, which decreases at $t=t_f$ with respect to its initial value. Finally, Fig.~\ref{fig:controle_thermal}(d) shows the obtained  control pulse. Note from Eq~(\ref{eq:A(t)}) that the pulse envelope does not have a bell-shape since the derivative of $\Phi(t)$ does not approach zero for $t\rightarrow0$ and $t\rightarrow t_f$. 

\section{Conclusion}

We derived an analytical formula for the control field capable of the simultaneous control of the dynamics of the populations and coherence phase of an open two-level system. The control is engineered by choosing a pair of dynamical function, $P(t)$ and $\Phi(t)$ which sets the dynamics of the population and the coherence phase, respectively, from a given initial condition to the desired final state. Given these functions, the absolute value of the coherence $C(t)$ can be determined for specific functions $P(t)$. By analyzing the equation for $C(t)$, the accessible transitions in population can be determined. It it worth noting that the accessible regions do not change essentially for different functional forms of $P(t)$. We have illustrated the control using several situations, showing that the control field is indeed successful in performing the desired population transitions and phase changes even in the presence of dephasing and thermal noise, once the transition is in the accessible region and the RWA can be considered. We also show that the envelope of the control field can be shaped by choosing appropriately the dynamical functions.

\pagebreak
\appendix
\section{Integrals for the coherence modulus}

Here we furnish explicit expressions for the integrals (\ref{eq:I1}) and  (\ref{eq:I2}) when the population dynamics $P(t)$ is chosen to be a sine squared function (\ref{eq:P(t)_deph}). The $I_1$ integral is given by,

\begin{align}
     I_1 (\tilde\Gamma,t)  = \pi f_1(t) + \left( \frac{P_f-P_i}{2} \right)
      \left[ f_1(t) -  f_2(t)\right],
\end{align}
where the auxiliary functions $f_1(t)$ and $f_2(t)$ are given by,
\begin{equation}
    f_{1}(t) = \frac{e^{2\tilde\Gamma t}-1}{2\tilde\Gamma}.
\end{equation}
and
\begin{equation}
    f_{2}(t) = e^{2\tilde\Gamma t}\left[ \frac{2\tilde\Gamma \cos \left( \frac{\pi}{t_f}t\right) + \frac{\pi}{t_f} \sin \left( \frac{\pi}{t_f}t \right) - 2\tilde\Gamma}{4\tilde\Gamma^2 + \frac{\pi^2}{t_f^2}}  \right]
\end{equation}

The $I_2$ integral can be written as a sum of three terms,
\begin{equation}
    I_2(\tilde\Gamma,t) = P_i^2  f_{1}(t)+h_1(t)+h_2(t),
\end{equation}
which are expressed as
\begin{equation}
    h_1(t) = \left( P_f - P_i \right)^2 \left( \frac{3}{8} f_{1}(t) - \frac{1}{2} f_{2}(t) + \frac{1}{8}f_{3}(t)\right),
\end{equation}
\begin{equation}
    h_2(t) = P_i\left(P_f-P_i \right) \left( f_{1}(t) - f_{2}(t)\right),
\end{equation}
Finally, the auxiliary function $f_3$(t) is defined as,
\begin{equation}
    f_{3}(t) = e^{2\tilde\Gamma t}\left[ \frac{2\tilde\Gamma \cos \left( \frac{2\pi}{t_f}t\right) + \frac{2\pi}{t_f} \sin \left( \frac{2\pi}{t_f}t \right) - 2\tilde\Gamma}{4\tilde\Gamma^2 + \frac{4\pi^2}{t_f^2}}  \right].
\end{equation}

\pagebreak


\begin{thebibliography}{37}%
\makeatletter
\providecommand \@ifxundefined [1]{%
 \@ifx{#1\undefined}
}%
\providecommand \@ifnum [1]{%
 \ifnum #1\expandafter \@firstoftwo
 \else \expandafter \@secondoftwo
 \fi
}%
\providecommand \@ifx [1]{%
 \ifx #1\expandafter \@firstoftwo
 \else \expandafter \@secondoftwo
 \fi
}%
\providecommand \natexlab [1]{#1}%
\providecommand \enquote  [1]{``#1''}%
\providecommand \bibnamefont  [1]{#1}%
\providecommand \bibfnamefont [1]{#1}%
\providecommand \citenamefont [1]{#1}%
\providecommand \href@noop [0]{\@secondoftwo}%
\providecommand \href [0]{\begingroup \@sanitize@url \@href}%
\providecommand \@href[1]{\@@startlink{#1}\@@href}%
\providecommand \@@href[1]{\endgroup#1\@@endlink}%
\providecommand \@sanitize@url [0]{\catcode `\\12\catcode `\$12\catcode
  `\&12\catcode `\#12\catcode `\^12\catcode `\_12\catcode `\%12\relax}%
\providecommand \@@startlink[1]{}%
\providecommand \@@endlink[0]{}%
\providecommand \url  [0]{\begingroup\@sanitize@url \@url }%
\providecommand \@url [1]{\endgroup\@href {#1}{\urlprefix }}%
\providecommand \urlprefix  [0]{URL }%
\providecommand \Eprint [0]{\href }%
\providecommand \doibase [0]{https://doi.org/}%
\providecommand \selectlanguage [0]{\@gobble}%
\providecommand \bibinfo  [0]{\@secondoftwo}%
\providecommand \bibfield  [0]{\@secondoftwo}%
\providecommand \translation [1]{[#1]}%
\providecommand \BibitemOpen [0]{}%
\providecommand \bibitemStop [0]{}%
\providecommand \bibitemNoStop [0]{.\EOS\space}%
\providecommand \EOS [0]{\spacefactor3000\relax}%
\providecommand \BibitemShut  [1]{\csname bibitem#1\endcsname}%
\let\auto@bib@innerbib\@empty
\bibitem [{\citenamefont {Leggett}\ \emph {et~al.}(1987)\citenamefont
  {Leggett}, \citenamefont {Chakravarty}, \citenamefont {Dorsey}, \citenamefont
  {Fisher}, \citenamefont {Garg},\ and\ \citenamefont
  {Zwerger}}]{RevModPhys.59.1}%
  \BibitemOpen
  \bibfield  {author} {\bibinfo {author} {\bibfnamefont {A.~J.}\ \bibnamefont
  {Leggett}}, \bibinfo {author} {\bibfnamefont {S.}~\bibnamefont
  {Chakravarty}}, \bibinfo {author} {\bibfnamefont {A.~T.}\ \bibnamefont
  {Dorsey}}, \bibinfo {author} {\bibfnamefont {M.~P.~A.}\ \bibnamefont
  {Fisher}}, \bibinfo {author} {\bibfnamefont {A.}~\bibnamefont {Garg}},\ and\
  \bibinfo {author} {\bibfnamefont {W.}~\bibnamefont {Zwerger}},\ }\bibfield
  {title} {\bibinfo {title} {Dynamics of the dissipative two-state system},\
  }\href {https://doi.org/10.1103/RevModPhys.59.1} {\bibfield  {journal}
  {\bibinfo  {journal} {Rev. Mod. Phys.}\ }\textbf {\bibinfo {volume} {59}},\
  \bibinfo {pages} {1} (\bibinfo {year} {1987})}\BibitemShut {NoStop}%
\bibitem [{\citenamefont {Allen}\ and\ \citenamefont {Eberly}(1987)}]{eberly}%
  \BibitemOpen
  \bibfield  {author} {\bibinfo {author} {\bibfnamefont {L.}~\bibnamefont
  {Allen}}\ and\ \bibinfo {author} {\bibfnamefont {J.~H.}\ \bibnamefont
  {Eberly}},\ }\href@noop {} {\emph {\bibinfo {title} {Optical Resonance and
  Two-Level Atoms}}}\ (\bibinfo  {publisher} {Dover Publications},\ \bibinfo
  {year} {1987})\BibitemShut {NoStop}%
\bibitem [{\citenamefont {Petta}\ \emph {et~al.}(2005)\citenamefont {Petta},
  \citenamefont {Johnson}, \citenamefont {Taylor}, \citenamefont {Laird},
  \citenamefont {Yacoby}, \citenamefont {Lukin}, \citenamefont {Marcus},
  \citenamefont {Hanson},\ and\ \citenamefont
  {Gossard}}]{doi:10.1126/science.1116955}%
  \BibitemOpen
  \bibfield  {author} {\bibinfo {author} {\bibfnamefont {J.~R.}\ \bibnamefont
  {Petta}}, \bibinfo {author} {\bibfnamefont {A.~C.}\ \bibnamefont {Johnson}},
  \bibinfo {author} {\bibfnamefont {J.~M.}\ \bibnamefont {Taylor}}, \bibinfo
  {author} {\bibfnamefont {E.~A.}\ \bibnamefont {Laird}}, \bibinfo {author}
  {\bibfnamefont {A.}~\bibnamefont {Yacoby}}, \bibinfo {author} {\bibfnamefont
  {M.~D.}\ \bibnamefont {Lukin}}, \bibinfo {author} {\bibfnamefont {C.~M.}\
  \bibnamefont {Marcus}}, \bibinfo {author} {\bibfnamefont {M.~P.}\
  \bibnamefont {Hanson}},\ and\ \bibinfo {author} {\bibfnamefont {A.~C.}\
  \bibnamefont {Gossard}},\ }\bibfield  {title} {\bibinfo {title} {Coherent
  manipulation of coupled electron spins in semiconductor quantum dots},\
  }\href {https://doi.org/10.1126/science.1116955} {\bibfield  {journal}
  {\bibinfo  {journal} {Science}\ }\textbf {\bibinfo {volume} {309}},\ \bibinfo
  {pages} {2180} (\bibinfo {year} {2005})},\ \Eprint
  {https://arxiv.org/abs/https://www.science.org/doi/pdf/10.1126/science.1116955}
  {https://www.science.org/doi/pdf/10.1126/science.1116955} \BibitemShut
  {NoStop}%
\bibitem [{\citenamefont {Jirari}\ and\ \citenamefont
  {Rabitz}(2024)}]{PhysRevA.110.042618}%
  \BibitemOpen
  \bibfield  {author} {\bibinfo {author} {\bibfnamefont {H.}~\bibnamefont
  {Jirari}}\ and\ \bibinfo {author} {\bibfnamefont {H.}~\bibnamefont
  {Rabitz}},\ }\bibfield  {title} {\bibinfo {title} {Quantum optimal control of
  the driven dissipative two-level system},\ }\href
  {https://doi.org/10.1103/PhysRevA.110.042618} {\bibfield  {journal} {\bibinfo
   {journal} {Phys. Rev. A}\ }\textbf {\bibinfo {volume} {110}},\ \bibinfo
  {pages} {042618} (\bibinfo {year} {2024})}\BibitemShut {NoStop}%
\bibitem [{\citenamefont {Weiss}(2008)}]{weiss2008quantum}%
  \BibitemOpen
  \bibfield  {author} {\bibinfo {author} {\bibfnamefont {U.}~\bibnamefont
  {Weiss}},\ }\href {https://doi.org/10.1142/6738} {\emph {\bibinfo {title}
  {Quantum Dissipative Systems}}},\ \bibinfo {edition} {3rd}\ ed.,\ Series in
  Modern Condensed Matter Physics\ (\bibinfo  {publisher} {World Scientific},\
  \bibinfo {year} {2008})\BibitemShut {NoStop}%
\bibitem [{\citenamefont {Bartee}\ \emph {et~al.}(2025)\citenamefont {Bartee},
  \citenamefont {Gilbert}, \citenamefont {Zuo}, \citenamefont {Das},
  \citenamefont {Tanttu}, \citenamefont {Yang}, \citenamefont
  {Dumoulin~Stuyck}, \citenamefont {Pauka}, \citenamefont {Su}, \citenamefont
  {Lim}, \citenamefont {Serrano}, \citenamefont {Escott}, \citenamefont
  {Hudson}, \citenamefont {Itoh}, \citenamefont {Laucht}, \citenamefont
  {Dzurak},\ and\ \citenamefont {Reilly}}]{Bartee2025}%
  \BibitemOpen
  \bibfield  {author} {\bibinfo {author} {\bibfnamefont {S.~K.}\ \bibnamefont
  {Bartee}}, \bibinfo {author} {\bibfnamefont {W.}~\bibnamefont {Gilbert}},
  \bibinfo {author} {\bibfnamefont {K.}~\bibnamefont {Zuo}}, \bibinfo {author}
  {\bibfnamefont {K.}~\bibnamefont {Das}}, \bibinfo {author} {\bibfnamefont
  {T.}~\bibnamefont {Tanttu}}, \bibinfo {author} {\bibfnamefont {C.~H.}\
  \bibnamefont {Yang}}, \bibinfo {author} {\bibfnamefont {N.}~\bibnamefont
  {Dumoulin~Stuyck}}, \bibinfo {author} {\bibfnamefont {S.~J.}\ \bibnamefont
  {Pauka}}, \bibinfo {author} {\bibfnamefont {R.~Y.}\ \bibnamefont {Su}},
  \bibinfo {author} {\bibfnamefont {W.~H.}\ \bibnamefont {Lim}}, \bibinfo
  {author} {\bibfnamefont {S.}~\bibnamefont {Serrano}}, \bibinfo {author}
  {\bibfnamefont {C.~C.}\ \bibnamefont {Escott}}, \bibinfo {author}
  {\bibfnamefont {F.~E.}\ \bibnamefont {Hudson}}, \bibinfo {author}
  {\bibfnamefont {K.~M.}\ \bibnamefont {Itoh}}, \bibinfo {author}
  {\bibfnamefont {A.}~\bibnamefont {Laucht}}, \bibinfo {author} {\bibfnamefont
  {A.~S.}\ \bibnamefont {Dzurak}},\ and\ \bibinfo {author} {\bibfnamefont
  {D.~J.}\ \bibnamefont {Reilly}},\ }\bibfield  {title} {\bibinfo {title}
  {Spin-qubit control with a milli-kelvin cmos chip},\ }\href
  {https://doi.org/10.1038/s41586-025-09157-x} {\bibfield  {journal} {\bibinfo
  {journal} {Nature}\ }\textbf {\bibinfo {volume} {643}},\ \bibinfo {pages}
  {382} (\bibinfo {year} {2025})}\BibitemShut {NoStop}%
\bibitem [{\citenamefont {Murch}\ \emph {et~al.}(2013)\citenamefont {Murch},
  \citenamefont {Weber}, \citenamefont {Macklin},\ and\ \citenamefont
  {Siddiqi}}]{Murch2013}%
  \BibitemOpen
  \bibfield  {author} {\bibinfo {author} {\bibfnamefont {K.~W.}\ \bibnamefont
  {Murch}}, \bibinfo {author} {\bibfnamefont {S.~J.}\ \bibnamefont {Weber}},
  \bibinfo {author} {\bibfnamefont {C.}~\bibnamefont {Macklin}},\ and\ \bibinfo
  {author} {\bibfnamefont {I.}~\bibnamefont {Siddiqi}},\ }\bibfield  {title}
  {\bibinfo {title} {Observing single quantum trajectories of a superconducting
  quantum bit},\ }\href {https://doi.org/10.1038/nature12539} {\bibfield
  {journal} {\bibinfo  {journal} {Nature}\ }\textbf {\bibinfo {volume} {502}},\
  \bibinfo {pages} {211} (\bibinfo {year} {2013})}\BibitemShut {NoStop}%
\bibitem [{\citenamefont {Brif}\ \emph {et~al.}(2010)\citenamefont {Brif},
  \citenamefont {Chakrabarti},\ and\ \citenamefont {Rabitz}}]{Brif_2010}%
  \BibitemOpen
  \bibfield  {author} {\bibinfo {author} {\bibfnamefont {C.}~\bibnamefont
  {Brif}}, \bibinfo {author} {\bibfnamefont {R.}~\bibnamefont {Chakrabarti}},\
  and\ \bibinfo {author} {\bibfnamefont {H.}~\bibnamefont {Rabitz}},\
  }\bibfield  {title} {\bibinfo {title} {Control of quantum phenomena: past,
  present and future},\ }\href {https://doi.org/10.1088/1367-2630/12/7/075008}
  {\bibfield  {journal} {\bibinfo  {journal} {New Journal of Physics}\ }\textbf
  {\bibinfo {volume} {12}},\ \bibinfo {pages} {075008} (\bibinfo {year}
  {2010})}\BibitemShut {NoStop}%
\bibitem [{\citenamefont {Beltrani}\ \emph {et~al.}(2011)\citenamefont
  {Beltrani}, \citenamefont {Dominy}, \citenamefont {Ho},\ and\ \citenamefont
  {Rabitz}}]{10.1063/1.3589404}%
  \BibitemOpen
  \bibfield  {author} {\bibinfo {author} {\bibfnamefont {V.}~\bibnamefont
  {Beltrani}}, \bibinfo {author} {\bibfnamefont {J.}~\bibnamefont {Dominy}},
  \bibinfo {author} {\bibfnamefont {T.-S.}\ \bibnamefont {Ho}},\ and\ \bibinfo
  {author} {\bibfnamefont {H.}~\bibnamefont {Rabitz}},\ }\bibfield  {title}
  {\bibinfo {title} {Exploring the top and bottom of the quantum control
  landscape},\ }\href {https://doi.org/10.1063/1.3589404} {\bibfield  {journal}
  {\bibinfo  {journal} {The Journal of Chemical Physics}\ }\textbf {\bibinfo
  {volume} {134}},\ \bibinfo {pages} {194106} (\bibinfo {year} {2011})},\
  \Eprint
  {https://arxiv.org/abs/https://pubs.aip.org/aip/jcp/article-pdf/doi/10.1063/1.3589404/13818402/194106\_1\_online.pdf}
  {https://pubs.aip.org/aip/jcp/article-pdf/doi/10.1063/1.3589404/13818402/194106\_1\_online.pdf}
  \BibitemShut {NoStop}%
\bibitem [{\citenamefont {Schirmer}\ \emph {et~al.}(2003)\citenamefont
  {Schirmer}, \citenamefont {Pullen},\ and\ \citenamefont
  {Solomon}}]{SCHIRMER2003281}%
  \BibitemOpen
  \bibfield  {author} {\bibinfo {author} {\bibfnamefont {S.~G.}\ \bibnamefont
  {Schirmer}}, \bibinfo {author} {\bibfnamefont {I.~C.}\ \bibnamefont
  {Pullen}},\ and\ \bibinfo {author} {\bibfnamefont {A.~I.}\ \bibnamefont
  {Solomon}},\ }\bibfield  {title} {\bibinfo {title} {Controllability of
  quantum systems},\ }\href
  {https://doi.org/https://doi.org/10.1016/S1474-6670(17)38905-X} {\bibfield
  {journal} {\bibinfo  {journal} {IFAC Proceedings Volumes}\ }\textbf {\bibinfo
  {volume} {36}},\ \bibinfo {pages} {281} (\bibinfo {year} {2003})},\ \bibinfo
  {note} {2nd IFAC Workshop on Lagrangian and Hamiltonian Methods for Nonlinear
  Control 2003, Seville, Spain, 3-5 April 2003}\BibitemShut {NoStop}%
\bibitem [{\citenamefont {Altafini}(2004)}]{PhysRevA.70.062321}%
  \BibitemOpen
  \bibfield  {author} {\bibinfo {author} {\bibfnamefont {C.}~\bibnamefont
  {Altafini}},\ }\bibfield  {title} {\bibinfo {title} {Coherent control of open
  quantum dynamical systems},\ }\href
  {https://doi.org/10.1103/PhysRevA.70.062321} {\bibfield  {journal} {\bibinfo
  {journal} {Phys. Rev. A}\ }\textbf {\bibinfo {volume} {70}},\ \bibinfo
  {pages} {062321} (\bibinfo {year} {2004})}\BibitemShut {NoStop}%
\bibitem [{\citenamefont {Yang}\ \emph {et~al.}(2013)\citenamefont {Yang},
  \citenamefont {Cong}, \citenamefont {Long}, \citenamefont {Ho}, \citenamefont
  {Wu},\ and\ \citenamefont {Rabitz}}]{PhysRevA.88.033420}%
  \BibitemOpen
  \bibfield  {author} {\bibinfo {author} {\bibfnamefont {F.}~\bibnamefont
  {Yang}}, \bibinfo {author} {\bibfnamefont {S.}~\bibnamefont {Cong}}, \bibinfo
  {author} {\bibfnamefont {R.}~\bibnamefont {Long}}, \bibinfo {author}
  {\bibfnamefont {T.-S.}\ \bibnamefont {Ho}}, \bibinfo {author} {\bibfnamefont
  {R.}~\bibnamefont {Wu}},\ and\ \bibinfo {author} {\bibfnamefont
  {H.}~\bibnamefont {Rabitz}},\ }\bibfield  {title} {\bibinfo {title}
  {Exploring the transition-probability-control landscape of open quantum
  systems: Application to a two-level case},\ }\href
  {https://doi.org/10.1103/PhysRevA.88.033420} {\bibfield  {journal} {\bibinfo
  {journal} {Phys. Rev. A}\ }\textbf {\bibinfo {volume} {88}},\ \bibinfo
  {pages} {033420} (\bibinfo {year} {2013})}\BibitemShut {NoStop}%
\bibitem [{\citenamefont {Koch}\ \emph {et~al.}(2022)\citenamefont {Koch},
  \citenamefont {Boscain}, \citenamefont {Calarco}, \citenamefont {Dirr},
  \citenamefont {Filipp}, \citenamefont {Glaser}, \citenamefont {Kosloff},
  \citenamefont {Montangero}, \citenamefont {Schulte-Herbr{\"u}ggen},
  \citenamefont {Sugny},\ and\ \citenamefont {Wilhelm}}]{Koch2022}%
  \BibitemOpen
  \bibfield  {author} {\bibinfo {author} {\bibfnamefont {C.~P.}\ \bibnamefont
  {Koch}}, \bibinfo {author} {\bibfnamefont {U.}~\bibnamefont {Boscain}},
  \bibinfo {author} {\bibfnamefont {T.}~\bibnamefont {Calarco}}, \bibinfo
  {author} {\bibfnamefont {G.}~\bibnamefont {Dirr}}, \bibinfo {author}
  {\bibfnamefont {S.}~\bibnamefont {Filipp}}, \bibinfo {author} {\bibfnamefont
  {S.~J.}\ \bibnamefont {Glaser}}, \bibinfo {author} {\bibfnamefont
  {R.}~\bibnamefont {Kosloff}}, \bibinfo {author} {\bibfnamefont
  {S.}~\bibnamefont {Montangero}}, \bibinfo {author} {\bibfnamefont
  {T.}~\bibnamefont {Schulte-Herbr{\"u}ggen}}, \bibinfo {author} {\bibfnamefont
  {D.}~\bibnamefont {Sugny}},\ and\ \bibinfo {author} {\bibfnamefont {F.~K.}\
  \bibnamefont {Wilhelm}},\ }\bibfield  {title} {\bibinfo {title} {Quantum
  optimal control in quantum technologies. strategic report on current status,
  visions and goals for research in europe},\ }\href
  {https://doi.org/10.1140/epjqt/s40507-022-00138-x} {\bibfield  {journal}
  {\bibinfo  {journal} {EPJ Quantum Technology}\ }\textbf {\bibinfo {volume}
  {9}},\ \bibinfo {pages} {19} (\bibinfo {year} {2022})}\BibitemShut {NoStop}%
\bibitem [{\citenamefont {Koch}(2016)}]{Koch_2016}%
  \BibitemOpen
  \bibfield  {author} {\bibinfo {author} {\bibfnamefont {C.~P.}\ \bibnamefont
  {Koch}},\ }\bibfield  {title} {\bibinfo {title} {Controlling open quantum
  systems: tools, achievements, and limitations},\ }\href
  {https://doi.org/10.1088/0953-8984/28/21/213001} {\bibfield  {journal}
  {\bibinfo  {journal} {Journal of Physics: Condensed Matter}\ }\textbf
  {\bibinfo {volume} {28}},\ \bibinfo {pages} {213001} (\bibinfo {year}
  {2016})}\BibitemShut {NoStop}%
\bibitem [{\citenamefont {Ansel}\ \emph {et~al.}(2024)\citenamefont {Ansel},
  \citenamefont {Dionis}, \citenamefont {Arrouas}, \citenamefont {Peaudecerf},
  \citenamefont {Guérin}, \citenamefont {Guéry-Odelin},\ and\ \citenamefont
  {Sugny}}]{Ansel_2024}%
  \BibitemOpen
  \bibfield  {author} {\bibinfo {author} {\bibfnamefont {Q.}~\bibnamefont
  {Ansel}}, \bibinfo {author} {\bibfnamefont {E.}~\bibnamefont {Dionis}},
  \bibinfo {author} {\bibfnamefont {F.}~\bibnamefont {Arrouas}}, \bibinfo
  {author} {\bibfnamefont {B.}~\bibnamefont {Peaudecerf}}, \bibinfo {author}
  {\bibfnamefont {S.}~\bibnamefont {Guérin}}, \bibinfo {author} {\bibfnamefont
  {D.}~\bibnamefont {Guéry-Odelin}},\ and\ \bibinfo {author} {\bibfnamefont
  {D.}~\bibnamefont {Sugny}},\ }\bibfield  {title} {\bibinfo {title}
  {Introduction to theoretical and experimental aspects of quantum optimal
  control},\ }\href {https://doi.org/10.1088/1361-6455/ad46a5} {\bibfield
  {journal} {\bibinfo  {journal} {Journal of Physics B: Atomic, Molecular and
  Optical Physics}\ }\textbf {\bibinfo {volume} {57}},\ \bibinfo {pages}
  {133001} (\bibinfo {year} {2024})}\BibitemShut {NoStop}%
\bibitem [{\citenamefont {Fernandes}\ \emph {et~al.}(2023)\citenamefont
  {Fernandes}, \citenamefont {Fanchini}, \citenamefont {de~Lima},\ and\
  \citenamefont {Castelano}}]{Fernandes_2023}%
  \BibitemOpen
  \bibfield  {author} {\bibinfo {author} {\bibfnamefont {M.~E.~F.}\
  \bibnamefont {Fernandes}}, \bibinfo {author} {\bibfnamefont {F.~F.}\
  \bibnamefont {Fanchini}}, \bibinfo {author} {\bibfnamefont {E.~F.}\
  \bibnamefont {de~Lima}},\ and\ \bibinfo {author} {\bibfnamefont {L.~K.}\
  \bibnamefont {Castelano}},\ }\bibfield  {title} {\bibinfo {title}
  {Effectiveness of the krotov method in finding controls for open quantum
  systems},\ }\href {https://doi.org/10.1088/1751-8121/ad0b5b} {\bibfield
  {journal} {\bibinfo  {journal} {Journal of Physics A: Mathematical and
  Theoretical}\ }\textbf {\bibinfo {volume} {56}},\ \bibinfo {pages} {495303}
  (\bibinfo {year} {2023})}\BibitemShut {NoStop}%
\bibitem [{\citenamefont {Shi}\ and\ \citenamefont
  {Rabitz}(1990)}]{10.1063/1.458438}%
  \BibitemOpen
  \bibfield  {author} {\bibinfo {author} {\bibfnamefont {S.}~\bibnamefont
  {Shi}}\ and\ \bibinfo {author} {\bibfnamefont {H.}~\bibnamefont {Rabitz}},\
  }\bibfield  {title} {\bibinfo {title} {{Quantum mechanical optimal control of
  physical observables in microsystems}},\ }\href
  {https://doi.org/10.1063/1.458438} {\bibfield  {journal} {\bibinfo  {journal}
  {The Journal of Chemical Physics}\ }\textbf {\bibinfo {volume} {92}},\
  \bibinfo {pages} {364} (\bibinfo {year} {1990})}\BibitemShut {NoStop}%
\bibitem [{\citenamefont {Dridi}\ \emph {et~al.}(2020)\citenamefont {Dridi},
  \citenamefont {Liu},\ and\ \citenamefont
  {Gu\'erin}}]{PhysRevLett.125.250403}%
  \BibitemOpen
  \bibfield  {author} {\bibinfo {author} {\bibfnamefont {G.}~\bibnamefont
  {Dridi}}, \bibinfo {author} {\bibfnamefont {K.}~\bibnamefont {Liu}},\ and\
  \bibinfo {author} {\bibfnamefont {S.}~\bibnamefont {Gu\'erin}},\ }\bibfield
  {title} {\bibinfo {title} {Optimal robust quantum control by inverse
  geometric optimization},\ }\href
  {https://doi.org/10.1103/PhysRevLett.125.250403} {\bibfield  {journal}
  {\bibinfo  {journal} {Phys. Rev. Lett.}\ }\textbf {\bibinfo {volume} {125}},\
  \bibinfo {pages} {250403} (\bibinfo {year} {2020})}\BibitemShut {NoStop}%
\bibitem [{\citenamefont {Shapiro}\ and\ \citenamefont
  {Brumer}(2011)}]{doi:https://doi.org/10.1002/9783527639700.ch5}%
  \BibitemOpen
  \bibfield  {author} {\bibinfo {author} {\bibfnamefont {M.}~\bibnamefont
  {Shapiro}}\ and\ \bibinfo {author} {\bibfnamefont {P.}~\bibnamefont
  {Brumer}},\ }\bibinfo {title} {Optimal control theory},\ in\ \href
  {https://doi.org/https://doi.org/10.1002/9783527639700.ch5} {\emph {\bibinfo
  {booktitle} {Quantum Control of Molecular Processes}}}\ (\bibinfo
  {publisher} {John Wiley \& Sons, Ltd},\ \bibinfo {year} {2011})\
  Chap.~\bibinfo {chapter} {5}, pp.\ \bibinfo {pages} {83--94}\BibitemShut
  {NoStop}%
\bibitem [{\citenamefont {Shen}\ and\ \citenamefont
  {Rabitz}(1994)}]{10.1063/1.467202}%
  \BibitemOpen
  \bibfield  {author} {\bibinfo {author} {\bibfnamefont {L.}~\bibnamefont
  {Shen}}\ and\ \bibinfo {author} {\bibfnamefont {H.}~\bibnamefont {Rabitz}},\
  }\bibfield  {title} {\bibinfo {title} {{Optimal control of vibronic
  population inversion with inclusion of molecular rotation}},\ }\href
  {https://doi.org/10.1063/1.467202} {\bibfield  {journal} {\bibinfo  {journal}
  {The Journal of Chemical Physics}\ }\textbf {\bibinfo {volume} {100}},\
  \bibinfo {pages} {4811} (\bibinfo {year} {1994})}\BibitemShut {NoStop}%
\bibitem [{\citenamefont {Mirrahimi}\ \emph {et~al.}(2005)\citenamefont
  {Mirrahimi}, \citenamefont {Turinici},\ and\ \citenamefont
  {Rouchon}}]{Mirrahimi2005}%
  \BibitemOpen
  \bibfield  {author} {\bibinfo {author} {\bibfnamefont {M.}~\bibnamefont
  {Mirrahimi}}, \bibinfo {author} {\bibfnamefont {G.}~\bibnamefont
  {Turinici}},\ and\ \bibinfo {author} {\bibfnamefont {P.}~\bibnamefont
  {Rouchon}},\ }\bibfield  {title} {\bibinfo {title} {Reference trajectory
  tracking for locally designed coherent quantum controls},\ }\href
  {https://doi.org/10.1021/jp0472461} {\bibfield  {journal} {\bibinfo
  {journal} {The Journal of Physical Chemistry A}\ }\textbf {\bibinfo {volume}
  {109}},\ \bibinfo {pages} {2631} (\bibinfo {year} {2005})}\BibitemShut
  {NoStop}%
\bibitem [{\citenamefont {Magann}\ \emph {et~al.}(2018)\citenamefont {Magann},
  \citenamefont {Ho},\ and\ \citenamefont {Rabitz}}]{PhysRevA.98.043429}%
  \BibitemOpen
  \bibfield  {author} {\bibinfo {author} {\bibfnamefont {A.}~\bibnamefont
  {Magann}}, \bibinfo {author} {\bibfnamefont {T.-S.}\ \bibnamefont {Ho}},\
  and\ \bibinfo {author} {\bibfnamefont {H.}~\bibnamefont {Rabitz}},\
  }\bibfield  {title} {\bibinfo {title} {Singularity-free quantum tracking
  control of molecular rotor orientation},\ }\href
  {https://doi.org/10.1103/PhysRevA.98.043429} {\bibfield  {journal} {\bibinfo
  {journal} {Phys. Rev. A}\ }\textbf {\bibinfo {volume} {98}},\ \bibinfo
  {pages} {043429} (\bibinfo {year} {2018})}\BibitemShut {NoStop}%
\bibitem [{\citenamefont {Chen}\ \emph {et~al.}(1997)\citenamefont {Chen},
  \citenamefont {Gross}, \citenamefont {Ramakrishna}, \citenamefont {Rabitz},
  \citenamefont {Mease},\ and\ \citenamefont {Singh}}]{CHEN19971617}%
  \BibitemOpen
  \bibfield  {author} {\bibinfo {author} {\bibfnamefont {Y.}~\bibnamefont
  {Chen}}, \bibinfo {author} {\bibfnamefont {P.}~\bibnamefont {Gross}},
  \bibinfo {author} {\bibfnamefont {V.}~\bibnamefont {Ramakrishna}}, \bibinfo
  {author} {\bibfnamefont {H.}~\bibnamefont {Rabitz}}, \bibinfo {author}
  {\bibfnamefont {K.}~\bibnamefont {Mease}},\ and\ \bibinfo {author}
  {\bibfnamefont {H.}~\bibnamefont {Singh}},\ }\bibfield  {title} {\bibinfo
  {title} {Control of classical regime molecular objectives—applications of
  tracking and variations on the theme},\ }\href
  {https://doi.org/https://doi.org/10.1016/S0005-1098(97)00077-0} {\bibfield
  {journal} {\bibinfo  {journal} {Automatica}\ }\textbf {\bibinfo {volume}
  {33}},\ \bibinfo {pages} {1617} (\bibinfo {year} {1997})}\BibitemShut
  {NoStop}%
\bibitem [{\citenamefont {Gross}\ \emph {et~al.}(1993)\citenamefont {Gross},
  \citenamefont {Singh}, \citenamefont {Rabitz}, \citenamefont {Mease},\ and\
  \citenamefont {Huang}}]{PhysRevA.47.4593}%
  \BibitemOpen
  \bibfield  {author} {\bibinfo {author} {\bibfnamefont {P.}~\bibnamefont
  {Gross}}, \bibinfo {author} {\bibfnamefont {H.}~\bibnamefont {Singh}},
  \bibinfo {author} {\bibfnamefont {H.}~\bibnamefont {Rabitz}}, \bibinfo
  {author} {\bibfnamefont {K.}~\bibnamefont {Mease}},\ and\ \bibinfo {author}
  {\bibfnamefont {G.~M.}\ \bibnamefont {Huang}},\ }\bibfield  {title} {\bibinfo
  {title} {Inverse quantum-mechanical control: A means for design and a test of
  intuition},\ }\href {https://doi.org/10.1103/PhysRevA.47.4593} {\bibfield
  {journal} {\bibinfo  {journal} {Phys. Rev. A}\ }\textbf {\bibinfo {volume}
  {47}},\ \bibinfo {pages} {4593} (\bibinfo {year} {1993})}\BibitemShut
  {NoStop}%
\bibitem [{\citenamefont {Magann}\ \emph {et~al.}(2023)\citenamefont {Magann},
  \citenamefont {Ho}, \citenamefont {Arenz},\ and\ \citenamefont
  {Rabitz}}]{PhysRevA.108.033106}%
  \BibitemOpen
  \bibfield  {author} {\bibinfo {author} {\bibfnamefont {A.~B.}\ \bibnamefont
  {Magann}}, \bibinfo {author} {\bibfnamefont {T.-S.}\ \bibnamefont {Ho}},
  \bibinfo {author} {\bibfnamefont {C.}~\bibnamefont {Arenz}},\ and\ \bibinfo
  {author} {\bibfnamefont {H.~A.}\ \bibnamefont {Rabitz}},\ }\bibfield  {title}
  {\bibinfo {title} {Quantum tracking control of the orientation of
  symmetric-top molecules},\ }\href
  {https://doi.org/10.1103/PhysRevA.108.033106} {\bibfield  {journal} {\bibinfo
   {journal} {Phys. Rev. A}\ }\textbf {\bibinfo {volume} {108}},\ \bibinfo
  {pages} {033106} (\bibinfo {year} {2023})}\BibitemShut {NoStop}%
\bibitem [{\citenamefont {Zhang}\ \emph {et~al.}(2017)\citenamefont {Zhang},
  \citenamefont {Chen},\ and\ \citenamefont {Gu{\'e}ry-Odelin}}]{Zhang2017}%
  \BibitemOpen
  \bibfield  {author} {\bibinfo {author} {\bibfnamefont {Q.}~\bibnamefont
  {Zhang}}, \bibinfo {author} {\bibfnamefont {X.}~\bibnamefont {Chen}},\ and\
  \bibinfo {author} {\bibfnamefont {D.}~\bibnamefont {Gu{\'e}ry-Odelin}},\
  }\bibfield  {title} {\bibinfo {title} {Reverse engineering protocols for
  controlling spin dynamics},\ }\href
  {https://doi.org/10.1038/s41598-017-16146-2} {\bibfield  {journal} {\bibinfo
  {journal} {Scientific Reports}\ }\textbf {\bibinfo {volume} {7}},\ \bibinfo
  {pages} {15814} (\bibinfo {year} {2017})}\BibitemShut {NoStop}%
\bibitem [{\citenamefont {Gonz\'alez-Resines}\ \emph
  {et~al.}(2017)\citenamefont {Gonz\'alez-Resines}, \citenamefont
  {Gu\'ery-Odelin}, \citenamefont {Tobalina}, \citenamefont {Lizuain},
  \citenamefont {Torrontegui},\ and\ \citenamefont
  {Muga}}]{PhysRevApplied.8.054008}%
  \BibitemOpen
  \bibfield  {author} {\bibinfo {author} {\bibfnamefont {S.}~\bibnamefont
  {Gonz\'alez-Resines}}, \bibinfo {author} {\bibfnamefont {D.}~\bibnamefont
  {Gu\'ery-Odelin}}, \bibinfo {author} {\bibfnamefont {A.}~\bibnamefont
  {Tobalina}}, \bibinfo {author} {\bibfnamefont {I.}~\bibnamefont {Lizuain}},
  \bibinfo {author} {\bibfnamefont {E.}~\bibnamefont {Torrontegui}},\ and\
  \bibinfo {author} {\bibfnamefont {J.~G.}\ \bibnamefont {Muga}},\ }\bibfield
  {title} {\bibinfo {title} {Invariant-based inverse engineering of crane
  control parameters},\ }\href
  {https://doi.org/10.1103/PhysRevApplied.8.054008} {\bibfield  {journal}
  {\bibinfo  {journal} {Phys. Rev. Appl.}\ }\textbf {\bibinfo {volume} {8}},\
  \bibinfo {pages} {054008} (\bibinfo {year} {2017})}\BibitemShut {NoStop}%
\bibitem [{\citenamefont {Wang}\ \emph {et~al.}(2024)\citenamefont {Wang},
  \citenamefont {Ma},\ and\ \citenamefont {Wu}}]{Wang2024}%
  \BibitemOpen
  \bibfield  {author} {\bibinfo {author} {\bibfnamefont {M.~Z.}\ \bibnamefont
  {Wang}}, \bibinfo {author} {\bibfnamefont {W.}~\bibnamefont {Ma}},\ and\
  \bibinfo {author} {\bibfnamefont {S.~L.}\ \bibnamefont {Wu}},\ }\bibfield
  {title} {\bibinfo {title} {Steady state engineering of a two-level system by
  the mixed-state inverse engineering scheme},\ }\href
  {https://doi.org/10.1038/s41598-024-53726-5} {\bibfield  {journal} {\bibinfo
  {journal} {Scientific Reports}\ }\textbf {\bibinfo {volume} {14}},\ \bibinfo
  {pages} {3409} (\bibinfo {year} {2024})}\BibitemShut {NoStop}%
\bibitem [{\citenamefont {Vitanov}\ and\ \citenamefont
  {Shore}(2015)}]{Vitanov_2015}%
  \BibitemOpen
  \bibfield  {author} {\bibinfo {author} {\bibfnamefont {N.~V.}\ \bibnamefont
  {Vitanov}}\ and\ \bibinfo {author} {\bibfnamefont {B.~W.}\ \bibnamefont
  {Shore}},\ }\bibfield  {title} {\bibinfo {title} {Designer evolution of
  quantum systems by inverse engineering},\ }\href
  {https://doi.org/10.1088/0953-4075/48/17/174008} {\bibfield  {journal}
  {\bibinfo  {journal} {Journal of Physics B: Atomic, Molecular and Optical
  Physics}\ }\textbf {\bibinfo {volume} {48}},\ \bibinfo {pages} {174008}
  (\bibinfo {year} {2015})}\BibitemShut {NoStop}%
\bibitem [{\citenamefont {Golubev}\ and\ \citenamefont
  {Kuleff}(2014)}]{articleGolubev}%
  \BibitemOpen
  \bibfield  {author} {\bibinfo {author} {\bibfnamefont {N.~V.}\ \bibnamefont
  {Golubev}}\ and\ \bibinfo {author} {\bibfnamefont {A.~I.}\ \bibnamefont
  {Kuleff}},\ }\bibfield  {title} {\bibinfo {title} {Control of populations of
  two-level systems by a single resonant laser pulse},\ }\href
  {https://doi.org/10.1103/PhysRevA.90.035401} {\bibfield  {journal} {\bibinfo
  {journal} {Phys. Rev. A}\ }\textbf {\bibinfo {volume} {90}},\ \bibinfo
  {pages} {035401} (\bibinfo {year} {2014})}\BibitemShut {NoStop}%
\bibitem [{\citenamefont {Csehi}(2019)}]{articleAndras}%
  \BibitemOpen
  \bibfield  {author} {\bibinfo {author} {\bibfnamefont {A.}~\bibnamefont
  {Csehi}},\ }\bibfield  {title} {\bibinfo {title} {Control of the populations
  and phases of two-level quantum systems by a single frequency-chirped laser
  pulse},\ }\href {https://doi.org/10.1088/1361-6455/ab3c05} {\bibfield
  {journal} {\bibinfo  {journal} {Journal of Physics B: Atomic, Molecular and
  Optical Physics}\ }\textbf {\bibinfo {volume} {52}},\ \bibinfo {pages}
  {195004} (\bibinfo {year} {2019})}\BibitemShut {NoStop}%
\bibitem [{\citenamefont {Fagundes}\ and\ \citenamefont
  {de~Lima}(2024)}]{PhysRevA.110.022201}%
  \BibitemOpen
  \bibfield  {author} {\bibinfo {author} {\bibfnamefont {F.~S.}\ \bibnamefont
  {Fagundes}}\ and\ \bibinfo {author} {\bibfnamefont {E.~F.}\ \bibnamefont
  {de~Lima}},\ }\bibfield  {title} {\bibinfo {title} {Reverse engineering
  control of the relative phase and populations of two-level quantum systems},\
  }\href {https://doi.org/10.1103/PhysRevA.110.022201} {\bibfield  {journal}
  {\bibinfo  {journal} {Phys. Rev. A}\ }\textbf {\bibinfo {volume} {110}},\
  \bibinfo {pages} {022201} (\bibinfo {year} {2024})}\BibitemShut {NoStop}%
\bibitem [{\citenamefont {Ran}\ \emph {et~al.}(2020{\natexlab{a}})\citenamefont
  {Ran}, \citenamefont {Zhang}, \citenamefont {Chen}, \citenamefont {Shi},
  \citenamefont {Xia}, \citenamefont {Ianconescu}, \citenamefont {Scheuer},\
  and\ \citenamefont {Gover}}]{Ran:20}%
  \BibitemOpen
  \bibfield  {author} {\bibinfo {author} {\bibfnamefont {D.}~\bibnamefont
  {Ran}}, \bibinfo {author} {\bibfnamefont {B.}~\bibnamefont {Zhang}}, \bibinfo
  {author} {\bibfnamefont {Y.-H.}\ \bibnamefont {Chen}}, \bibinfo {author}
  {\bibfnamefont {Z.-C.}\ \bibnamefont {Shi}}, \bibinfo {author} {\bibfnamefont
  {Y.}~\bibnamefont {Xia}}, \bibinfo {author} {\bibfnamefont {R.}~\bibnamefont
  {Ianconescu}}, \bibinfo {author} {\bibfnamefont {J.}~\bibnamefont
  {Scheuer}},\ and\ \bibinfo {author} {\bibfnamefont {A.}~\bibnamefont
  {Gover}},\ }\bibfield  {title} {\bibinfo {title} {Effective pulse
  reverse-engineering for strong field--matter interaction},\ }\href
  {https://doi.org/10.1364/OL.397053} {\bibfield  {journal} {\bibinfo
  {journal} {Opt. Lett.}\ }\textbf {\bibinfo {volume} {45}},\ \bibinfo {pages}
  {3597} (\bibinfo {year} {2020}{\natexlab{a}})}\BibitemShut {NoStop}%
\bibitem [{\citenamefont {Xiao}\ \emph {et~al.}(2021)\citenamefont {Xiao},
  \citenamefont {Ke},\ and\ \citenamefont {Ji}}]{XIAO2021167957}%
  \BibitemOpen
  \bibfield  {author} {\bibinfo {author} {\bibfnamefont {S.}~\bibnamefont
  {Xiao}}, \bibinfo {author} {\bibfnamefont {Q.}~\bibnamefont {Ke}},\ and\
  \bibinfo {author} {\bibfnamefont {Y.}~\bibnamefont {Ji}},\ }\bibfield
  {title} {\bibinfo {title} {Research on reverse engineering strategy of state
  transfer for open quantum systems},\ }\href
  {https://doi.org/https://doi.org/10.1016/j.ijleo.2021.167957} {\bibfield
  {journal} {\bibinfo  {journal} {Optik}\ }\textbf {\bibinfo {volume} {247}},\
  \bibinfo {pages} {167957} (\bibinfo {year} {2021})}\BibitemShut {NoStop}%
\bibitem [{\citenamefont {Medina}\ and\ \citenamefont
  {Semi\~ao}(2019)}]{PhysRevA.100.012103}%
  \BibitemOpen
  \bibfield  {author} {\bibinfo {author} {\bibfnamefont {I.}~\bibnamefont
  {Medina}}\ and\ \bibinfo {author} {\bibfnamefont {F.~L.}\ \bibnamefont
  {Semi\~ao}},\ }\bibfield  {title} {\bibinfo {title} {Pulse engineering for
  population control under dephasing and dissipation},\ }\href
  {https://doi.org/10.1103/PhysRevA.100.012103} {\bibfield  {journal} {\bibinfo
   {journal} {Phys. Rev. A}\ }\textbf {\bibinfo {volume} {100}},\ \bibinfo
  {pages} {012103} (\bibinfo {year} {2019})}\BibitemShut {NoStop}%
\bibitem [{\citenamefont {Ran}\ \emph {et~al.}(2020{\natexlab{b}})\citenamefont
  {Ran}, \citenamefont {Shan}, \citenamefont {Shi}, \citenamefont {Yang},
  \citenamefont {Song},\ and\ \citenamefont {Xia}}]{PhysRevA.101.023822}%
  \BibitemOpen
  \bibfield  {author} {\bibinfo {author} {\bibfnamefont {D.}~\bibnamefont
  {Ran}}, \bibinfo {author} {\bibfnamefont {W.-J.}\ \bibnamefont {Shan}},
  \bibinfo {author} {\bibfnamefont {Z.-C.}\ \bibnamefont {Shi}}, \bibinfo
  {author} {\bibfnamefont {Z.-B.}\ \bibnamefont {Yang}}, \bibinfo {author}
  {\bibfnamefont {J.}~\bibnamefont {Song}},\ and\ \bibinfo {author}
  {\bibfnamefont {Y.}~\bibnamefont {Xia}},\ }\bibfield  {title} {\bibinfo
  {title} {Pulse reverse engineering for controlling two-level quantum
  systems},\ }\href {https://doi.org/10.1103/PhysRevA.101.023822} {\bibfield
  {journal} {\bibinfo  {journal} {Phys. Rev. A}\ }\textbf {\bibinfo {volume}
  {101}},\ \bibinfo {pages} {023822} (\bibinfo {year}
  {2020}{\natexlab{b}})}\BibitemShut {NoStop}%
\bibitem [{\citenamefont {Lünnemann}\ \emph {et~al.}(2008)\citenamefont
  {Lünnemann}, \citenamefont {Kuleff},\ and\ \citenamefont
  {Cederbaum}}]{10.1063/1.2970088}%
  \BibitemOpen
  \bibfield  {author} {\bibinfo {author} {\bibfnamefont {S.}~\bibnamefont
  {Lünnemann}}, \bibinfo {author} {\bibfnamefont {A.~I.}\ \bibnamefont
  {Kuleff}},\ and\ \bibinfo {author} {\bibfnamefont {L.~S.}\ \bibnamefont
  {Cederbaum}},\ }\bibfield  {title} {\bibinfo {title} {Charge migration
  following ionization in systems with chromophore-donor and amine-acceptor
  sites},\ }\href {https://doi.org/10.1063/1.2970088} {\bibfield  {journal}
  {\bibinfo  {journal} {The Journal of Chemical Physics}\ }\textbf {\bibinfo
  {volume} {129}},\ \bibinfo {pages} {104305} (\bibinfo {year} {2008})},\
  \Eprint
  {https://arxiv.org/abs/https://pubs.aip.org/aip/jcp/article-pdf/doi/10.1063/1.2970088/11067631/104305\_1\_online.pdf}
  {https://pubs.aip.org/aip/jcp/article-pdf/doi/10.1063/1.2970088/11067631/104305\_1\_online.pdf}
  \BibitemShut {NoStop}%
\end{thebibliography}

%

\end{document}